\pdfoutput=1
\documentclass[11pt,a4paper]{article}
\usepackage[utf8]{inputenc}
\usepackage[T1]{fontenc}
\usepackage{amsmath,amssymb,amsthm}
\usepackage{graphicx,microtype,lmodern}
\usepackage{pgfplots}
\usepgfplotslibrary{groupplots}
\pgfplotsset{compat=1.18}
\usepackage{booktabs}
\usepackage[margin=2.3cm]{geometry}
\usepackage{xcolor}
\usepackage{hyperref}
\hypersetup{colorlinks=true,linkcolor=blue!60!black,citecolor=blue!60!black,urlcolor=blue!60!black}

\newtheorem{proposition}{Proposition}
\newtheorem{lemma}{Lemma}
\newcommand{\sgn}{\operatorname{sgn}}
\newcommand{\Tr}{\operatorname{Tr}}
\newcommand{\ket}[1]{|#1\rangle}
\newcommand{\bra}[1]{\langle#1|}

\newcommand{\Zt}{\widetilde Z}
\newcommand{\Yt}{\widetilde Y}

\title{Discrete quadrature spectrum of composite bosons:\\ Weyl relation, intensity correlations, and binned homodyne detection}
\author{I.~Roditi\\[4pt]
\small CBPF -- Centro Brasileiro de Pesquisas F\'isicas\\
\small Rua Dr.~Xavier Sigaud 150, 22290-180, Rio de Janeiro, Brazil\\
\small \href{mailto:roditi@cbpf.br}{roditi@cbpf.br}}
\date{4 November 2026}

\begin{document}
\maketitle

\begin{abstract}
A composite boson made of two fermions can look elementary under one measurement and composite under another. In an exactly solvable paired-mode model we compare three probes: classical drives, periodically binned quadrature statistics, and intensity correlations $g^{(2)}$. The Weyl defect has an exact product formula and a controlled dilute expansion in the Schmidt moments. The binned observables are instead piecewise constant in the Schmidt weights and sensitive to single eigenvalues crossing bin edges; on the flat-spectrum ladder they converge slowly and nonmonotonically over the computed range. A spectral filter and a polar correction recover an exact Pauli algebra; a heralded joint projection then extracts a logical singlet from the vacua of two independent copies of the composite, with probability set by the accepted weight and the mixedness of the filtered reduced qubit. Finite-mode examples with Schmidt number about three succeed at the percent level despite order-one Weyl defects; near-unit $g^{(2)}$ does not control the binned algebra. For flat spectra the construction converges strongly to its oscillator counterpart, whose vacuum overlap is strictly positive at every scale; for infinite spectra with summable Schmidt amplitudes it converges at all but countably many filter cutoffs, positivity at the cutoffs studied remaining open.
\end{abstract}

\section{Introduction}

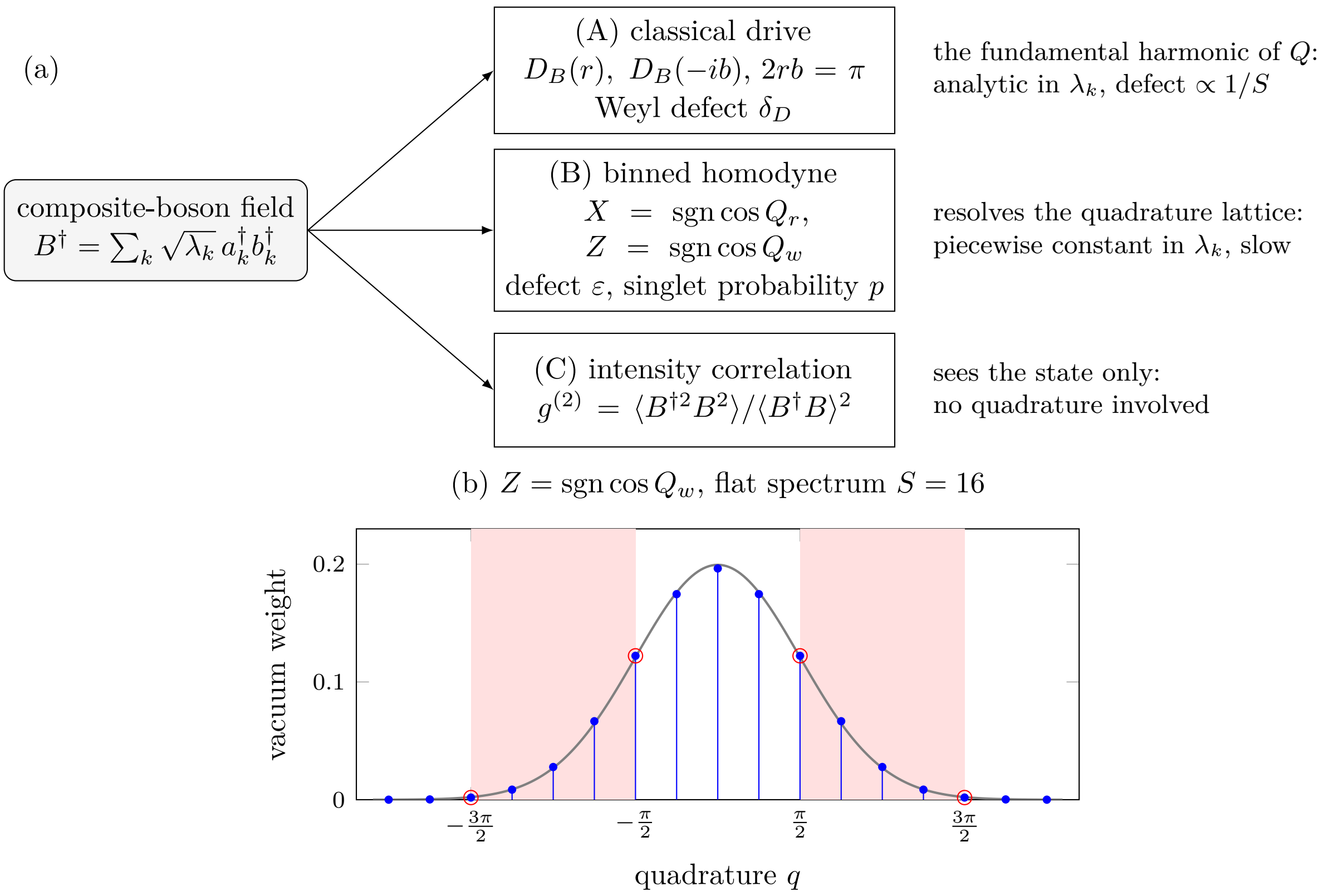
\begin{figure}[t]
\centering
\begin{tikzpicture}[font=\small,>=latex]
\node[draw,rounded corners,align=center,minimum width=3.6cm,minimum height=1.2cm,fill=gray!8] (F) at (0,0) {composite-boson field\\[1pt] $B^\dagger=\sum_k\sqrt{\lambda_k}\,a_k^\dagger b_k^\dagger$};
\node[draw,align=center,text width=4.5cm,minimum height=1.35cm] (A) at (6.4,1.9) {(A) classical drive\\[1pt] $D_B(r),\ D_B(-ib)$,\ $2rb=\pi$\\[1pt] Weyl defect $\delta_D$};
\node[draw,align=center,text width=4.5cm,minimum height=1.35cm] (C) at (6.4,-1.9) {(C) intensity correlation\\[1pt] $g^{(2)}=\langle B^{\dagger2}B^2\rangle/\langle B^\dagger B\rangle^2$};
\node[draw,align=center,text width=4.5cm,minimum height=1.35cm] (B) at (6.4,0) {(B) binned homodyne\\[1pt] $X=\sgn\cos Q_r$,\ $Z=\sgn\cos Q_w$\\[1pt] defect $\varepsilon$, singlet probability $p$};
\draw[->] (F.east) -- (A.west); \draw[->] (F.east) -- (B.west); \draw[->] (F.east) -- (C.west);
\node[align=left,font=\footnotesize,anchor=west] at (9.1,1.9) {the fundamental harmonic of $Q$:\\ analytic in $\lambda_k$, defect $\propto1/S$};
\node[align=left,font=\footnotesize,anchor=west] at (9.1,0) {resolves the quadrature lattice:\\ piecewise constant in $\lambda_k$, slow};
\node[align=left,font=\footnotesize,anchor=west] at (9.1,-1.9) {sees the state only:\\ no quadrature involved};
\node[font=\small,anchor=east] at (-1.0,1.9) {(a)};
\end{tikzpicture}

\vspace{3pt}

\begin{tikzpicture}
\begin{axis}[width=0.62\textwidth,height=4.8cm,xmin=-6.9,xmax=6.9,ymin=0,ymax=0.23,xlabel={quadrature $q$},ylabel={vacuum weight},xtick={-4.712,-1.571,1.571,4.712},xticklabels={$-\tfrac{3\pi}{2}$,$-\tfrac{\pi}{2}$,$\tfrac{\pi}{2}$,$\tfrac{3\pi}{2}$},tick label style={font=\scriptsize},label style={font=\small},title={(b) $Z=\sgn\cos Q_w$, flat spectrum $S=16$},title style={font=\small}]
\addplot[fill=red!12,draw=none] coordinates {(-4.712,0) (-4.712,0.23) (-1.571,0.23) (-1.571,0)} \closedcycle;
\addplot[fill=red!12,draw=none] coordinates {(1.571,0) (1.571,0.23) (4.712,0.23) (4.712,0)} \closedcycle;
\addplot[gray,thick] coordinates {(-6.5832,0.00003) (-6.5283,0.00004) (-6.4735,0.00004) (-6.4186,0.00005) (-6.3637,0.00005) (-6.3089,0.00006) (-6.2540,0.00007) (-6.1992,0.00008) (-6.1443,0.00009) (-6.0894,0.00011) (-6.0346,0.00012) (-5.9797,0.00014) (-5.9249,0.00016) (-5.8700,0.00019) (-5.8151,0.00021) (-5.7603,0.00024) (-5.7054,0.00027) (-5.6506,0.00031) (-5.5957,0.00035) (-5.5408,0.00040) (-5.4860,0.00045) (-5.4311,0.00051) (-5.3763,0.00057) (-5.3214,0.00064) (-5.2665,0.00072) (-5.2117,0.00081) (-5.1568,0.00091) (-5.1020,0.00102) (-5.0471,0.00114) (-4.9922,0.00128) (-4.9374,0.00143) (-4.8825,0.00159) (-4.8277,0.00177) (-4.7728,0.00197) (-4.7179,0.00219) (-4.6631,0.00243) (-4.6082,0.00270) (-4.5534,0.00299) (-4.4985,0.00330) (-4.4437,0.00365) (-4.3888,0.00402) (-4.3339,0.00443) (-4.2791,0.00488) (-4.2242,0.00536) (-4.1694,0.00589) (-4.1145,0.00646) (-4.0596,0.00707) (-4.0048,0.00773) (-3.9499,0.00845) (-3.8951,0.00922) (-3.8402,0.01005) (-3.7853,0.01094) (-3.7305,0.01189) (-3.6756,0.01291) (-3.6208,0.01400) (-3.5659,0.01516) (-3.5110,0.01641) (-3.4562,0.01773) (-3.4013,0.01913) (-3.3465,0.02062) (-3.2916,0.02220) (-3.2367,0.02387) (-3.1819,0.02564) (-3.1270,0.02750) (-3.0722,0.02946) (-3.0173,0.03153) (-2.9624,0.03369) (-2.9076,0.03596) (-2.8527,0.03834) (-2.7979,0.04083) (-2.7430,0.04342) (-2.6881,0.04613) (-2.6333,0.04894) (-2.5784,0.05186) (-2.5236,0.05488) (-2.4687,0.05801) (-2.4138,0.06125) (-2.3590,0.06459) (-2.3041,0.06802) (-2.2493,0.07156) (-2.1944,0.07518) (-2.1395,0.07889) (-2.0847,0.08268) (-2.0298,0.08655) (-1.9750,0.09049) (-1.9201,0.09450) (-1.8652,0.09856) (-1.8104,0.10267) (-1.7555,0.10682) (-1.7007,0.11101) (-1.6458,0.11521) (-1.5909,0.11943) (-1.5361,0.12366) (-1.4812,0.12788) (-1.4264,0.13208) (-1.3715,0.13625) (-1.3166,0.14038) (-1.2618,0.14447) (-1.2069,0.14849) (-1.1521,0.15243) (-1.0972,0.15629) (-1.0423,0.16005) (-0.9875,0.16371) (-0.9326,0.16724) (-0.8778,0.17064) (-0.8229,0.17389) (-0.7680,0.17700) (-0.7132,0.17994) (-0.6583,0.18270) (-0.6035,0.18528) (-0.5486,0.18767) (-0.4937,0.18986) (-0.4389,0.19184) (-0.3840,0.19360) (-0.3292,0.19514) (-0.2743,0.19645) (-0.2194,0.19753) (-0.1646,0.19838) (-0.1097,0.19899) (-0.0549,0.19935) (0.0000,0.19947) (0.0549,0.19935) (0.1097,0.19899) (0.1646,0.19838) (0.2194,0.19753) (0.2743,0.19645) (0.3292,0.19514) (0.3840,0.19360) (0.4389,0.19184) (0.4937,0.18986) (0.5486,0.18767) (0.6035,0.18528) (0.6583,0.18270) (0.7132,0.17994) (0.7680,0.17700) (0.8229,0.17389) (0.8778,0.17064) (0.9326,0.16724) (0.9875,0.16371) (1.0423,0.16005) (1.0972,0.15629) (1.1521,0.15243) (1.2069,0.14849) (1.2618,0.14447) (1.3166,0.14038) (1.3715,0.13625) (1.4264,0.13208) (1.4812,0.12788) (1.5361,0.12366) (1.5909,0.11943) (1.6458,0.11521) (1.7007,0.11101) (1.7555,0.10682) (1.8104,0.10267) (1.8652,0.09856) (1.9201,0.09450) (1.9750,0.09049) (2.0298,0.08655) (2.0847,0.08268) (2.1395,0.07889) (2.1944,0.07518) (2.2493,0.07156) (2.3041,0.06802) (2.3590,0.06459) (2.4138,0.06125) (2.4687,0.05801) (2.5236,0.05488) (2.5784,0.05186) (2.6333,0.04894) (2.6881,0.04613) (2.7430,0.04342) (2.7979,0.04083) (2.8527,0.03834) (2.9076,0.03596) (2.9624,0.03369) (3.0173,0.03153) (3.0722,0.02946) (3.1270,0.02750) (3.1819,0.02564) (3.2367,0.02387) (3.2916,0.02220) (3.3465,0.02062) (3.4013,0.01913) (3.4562,0.01773) (3.5110,0.01641) (3.5659,0.01516) (3.6208,0.01400) (3.6756,0.01291) (3.7305,0.01189) (3.7853,0.01094) (3.8402,0.01005) (3.8951,0.00922) (3.9499,0.00845) (4.0048,0.00773) (4.0596,0.00707) (4.1145,0.00646) (4.1694,0.00589) (4.2242,0.00536) (4.2791,0.00488) (4.3339,0.00443) (4.3888,0.00402) (4.4437,0.00365) (4.4985,0.00330) (4.5534,0.00299) (4.6082,0.00270) (4.6631,0.00243) (4.7179,0.00219) (4.7728,0.00197) (4.8277,0.00177) (4.8825,0.00159) (4.9374,0.00143) (4.9922,0.00128) (5.0471,0.00114) (5.1020,0.00102) (5.1568,0.00091) (5.2117,0.00081) (5.2665,0.00072) (5.3214,0.00064) (5.3763,0.00057) (5.4311,0.00051) (5.4860,0.00045) (5.5408,0.00040) (5.5957,0.00035) (5.6506,0.00031) (5.7054,0.00027) (5.7603,0.00024) (5.8151,0.00021) (5.8700,0.00019) (5.9249,0.00016) (5.9797,0.00014) (6.0346,0.00012) (6.0894,0.00011) (6.1443,0.00009) (6.1992,0.00008) (6.2540,0.00007) (6.3089,0.00006) (6.3637,0.00005) (6.4186,0.00005) (6.4735,0.00004) (6.5283,0.00004) (6.5832,0.00003)};
\addplot[blue,ycomb,mark=*,mark size=1.2pt] coordinates {(6.28319,0.00002) (5.49779,0.00024) (4.71239,0.00183) (3.92699,0.00854) (3.14159,0.02777) (2.35619,0.06665) (1.57080,0.12219) (0.78540,0.17456) (0.00000,0.19638) (-0.78540,0.17456) (-1.57080,0.12219) (-2.35619,0.06665) (-3.14159,0.02777) (-3.92699,0.00854) (-4.71239,0.00183) (-5.49779,0.00024) (-6.28319,0.00002)};
\addplot[red,only marks,mark=o,mark size=2.4pt] coordinates {(4.71239,0.00183) (1.57080,0.12219) (-1.57080,0.12219) (-4.71239,0.00183)};
\end{axis}
\end{tikzpicture}
\caption{(a) The three detectors applied to one composite-boson field, and what each of them resolves of the quadrature spectrum. (b) The discrete quadrature spectrum behind the results: eigenvalues of $Q_w=-(\pi/2)(B+B^\dagger)$ for a Frenkel-like coboson with $S=16$ sites (stems, with the vacuum weight on each), against the bins of the sign function (shaded: $Z=-1$) and the oscillator vacuum distribution (grey line, scaled to the lattice spacing). The lattice has spacing $\pi/\sqrt S$ against bins of width $\pi$; here four lattice points fall exactly on bin edges (circles), carrying $25\%$ of the vacuum weight, so the sign of $Z$ on them is a convention and the binned operator at this rank depends on it (Sec.~\ref{sec:finite}). A drive responds to this distribution through its characteristic function at one frequency, a smooth functional dominated in the dilute regime by the first moments; binned homodyne sees every point.}
\label{fig:scheme}
\end{figure}

Excitons in semiconductors and molecular crystals, Feshbach molecules of cold atoms, and biphotons are pairs of fermions or of bosons that behave as bosons when the pairs are dilute and their constituents are not resolved. How bosonic such a composite is has been quantified in several ways: by the normalisation ratios $F_{N+1}/F_N$ of the $N$-pair states and the deviation of the commutator $[B,B^\dagger]$ from unity in the coboson many-body formalism \cite{Combescot2008,CombescotShiau2015,Shiau2017,Figueiredo2026}, by the Schmidt number or entanglement of the pair wavefunction \cite{Law2005,Chudzicki2010,Tichy2012,Tichy2014}, or by the second-order correlation function $g^{(2)}$ of coherent states \cite{Shiau2017}, which is also the quantity measured when exciton or polariton condensates are tested for coherence \cite{Kasprzak2006,Deng2010,Combescot2017}. All of these are properties of the state or of the field algebra. A measurement, however, asks a narrower question: can \emph{this} detector tell the composite from an elementary boson? The present paper answers that question for three detectors applied to the same composite field (Fig.~\ref{fig:scheme}), and finds that the answers are different.

The three detectors are the following. (A) A \emph{classical drive}: two displacements $D_B(r)$ and $D_B(-ib)$ of the composite field with $2rb=\pi$, which for an elementary boson anticommute exactly (the Weyl relation). (B) \emph{Homodyne detection with periodically binned outcomes}: the quadrature outcome $q$ is recorded only through the sign of $\cos q$, which converts the two conjugate quadratures into dichotomic observables $X=\sgn\cos Q_r$ and $Z=\sgn\cos Q_w$. For an elementary boson these two involutions (operators squaring to the identity) anticommute exactly, so that two independent oscillators -- one held by each of two parties, Alice and Bob -- carry commuting Pauli algebras, and their joint vacuum $\ket0_A\ket0_B$ (the ``two vacua'' referred to throughout) has a nonzero overlap with the spin singlet of the two Pauli pairs \cite{Sorella2026}. Dichotomised homodyne outcomes have a long history in continuous-variable Bell tests, from sign binning \cite{Gilchrist1998,Munro1999,Wenger2003,GarciaPatron2004} to parity and pseudospin operators \cite{Banaszek1999,Chen2002}; periodic binning is the modular-variable construction \cite{Aharonov1969,VernazGris2014,Ketterer2016}, and the same coarse graining underlies the mutually unbiased coarse-grained measurements of Refs.~\cite{Tasca2018,Paul2018}; Bell--CHSH tests built from Weyl operators and squeezed states have also been formulated directly for bosonic quantum fields \cite{Dudal2023}. In all of that work the field is an elementary boson; here it is a composite. (C) \emph{Intensity correlations}: the normally ordered moment $g^{(2)}=\langle B^{\dagger2}B^2\rangle/\langle B^\dagger B\rangle^2$ measured by photon or atom counting. Of the three, detector (B) is the one that matters if the composite is to be used as a qubit resource, and it is the one for which bosonicity has not been examined.

Detector (B) is not an arbitrary choice. Sign-binned quadratures at the Weyl area $2rb=\pi$ are the periodic coarse graining of the modular-variable constructions cited above, and the Gottesman--Kitaev--Preskill encoding \cite{GKP2001} is built on the same structure: its logical Pauli operators are displacements, and its logical measurement is a homodyne measurement whose outcome is decoded periodically. The globally anticommuting binned quadratures studied here are one realisation of a qubit algebra in a continuous-variable field -- not the only one, since a qubit can also be encoded in two oscillator states -- but it is the one that these constructions use, and all of them have been formulated for elementary bosons. If composite fields -- exciton-polaritons, molecular condensates, or the collective excitations of spin ensembles -- are to serve in such protocols, the relevant question is whether \emph{their} binned quadratures anticommute, and the conventional bosonicity criteria do not answer it: Secs.~\ref{sec:finite} and \ref{sec:initial} show that a Weyl defect of $10^{-3}$ coexists with a binned defect of $0.37$, and $g^{(2)}=0.99999$ with a binned defect of order one. The question has a positive side as well: Sec.~\ref{sec:recovery} shows that a composite whose binned quadratures fail to anticommute can still yield an exact logical qubit, with the rejected weight computable from the spectrum.

Where such measurements exist. For a flat spectrum the composite field is a collective spin, $B=J_-/\sqrt S$ (Sec.~\ref{sec:model}), and the two binned quadratures are coarse-grained measurements of $J_x$ and $J_y$ on an ensemble of $S$ two-level systems -- the Holstein--Primakoff setting \cite{HolsteinPrimakoff1940}, in which collective-spin measurements are routine \cite{Ma2011}. Our flat-spectrum results are therefore exact statements about when the bosonic approximation of a finite ensemble survives a sign-binned measurement. For exciton-polaritons the quadratures of the emitted light are measured by homodyne detection \cite{Deng2010}, although relating them to the composite field requires a coupling model (Sec.~\ref{sec:model}); for trapped dimers the coboson formalism applies directly \cite{Jimenez2023}, but a quadrature measurement of a molecular field has no established implementation. In every case the binned measurement changes the pair number and therefore presupposes a phase reference or reservoir (Sec.~\ref{sec:singlet}), as does any displacement.

At finite Schmidt rank, the quadrature of the composite has a \emph{discrete spectrum}, which makes its response to binning sensitive to individual eigenvalues. If $B^\dagger=\sum_k\sqrt{\lambda_k}\,a_k^\dagger b_k^\dagger$ is the Schmidt decomposition of the pair, the eigenvalues of a quadrature are the signed sums $\sum_k\pm\sqrt{\lambda_k}$: $2^S$ values counted with multiplicity for $S$ Schmidt modes, which for a flat spectrum reduce to $S+1$ equally spaced points with spacing $\propto1/\sqrt S$, whereas for the infinite geometric family the support is an interval or a Cantor set depending on the decay parameter. For both infinite families considered below, the quadrature spectral measures have no atoms\footnote{An \emph{atom} of a measure is a single point carrying positive measure. The spectral measure of a quadrature $Q$ in a state $\psi$ assigns to each set of real numbers the probability that a measurement of $Q$ in $\psi$ gives a value in that set; it has no atoms when no single value has positive probability. The bin edges of $\sgn\cos$ then carry no weight, which is the property the convergence arguments of Sec.~\ref{sec:purity} use.} (Sec.~\ref{sec:purity}). An elementary boson's quadrature is continuous. The probes respond differently to this spectral structure. A drive is a smooth probe -- the fundamental harmonic of the quadrature -- so its defect is a real-analytic function of the Schmidt weights, computable in closed form, with a dilute expansion controlled by the first Schmidt moments. Intensity correlations do not involve the quadrature at all. Binned homodyne, with bins of fixed width against a lattice whose spacing shrinks with $S$, resolves the lattice point by point: its operators change only when a lattice point crosses a bin edge, they are therefore piecewise constant in the Schmidt weights, the purity and the third moment do not determine them, and their convergence to the oscillator algebra is slow and nonmonotone. Pauli blocking thus has two faces -- the commutator, seen by the drive, and the discreteness of the quadrature spectrum, seen by binned homodyne -- and neither the moment expansions of coboson theory nor $g^{(2)}$ bridges them.

Two conclusions follow. First, bosonic quality is not what the qubit task requires: a composite with Schmidt number $3.00$ (a $19$-mode truncation of a geometric spectrum), whose normalisation ratio $F_2/F_1=1-P$ is a third below unity and whose Weyl defect is of order one, still yields an exact qubit singlet from two vacua with probability $1.2\%$, after a spectral filter and a polar correction that we specify. Second, the singlet can be extracted from two vacua only because, in the factorisation that the two observables define, the qubit is entangled with its complementary logical factor -- a structure that the elementary oscillator shares and that compositeness modulates: the success probability is the squared accepted weight times the mixedness of the filtered reduced qubit, a function of the Schmidt spectrum which vanishes at many small ranks and vanishes identically when the spectrum is so concentrated that $\sum_k\sqrt{\lambda_k}<\pi/2$ at $r=1$, for then one binned quadrature is the identity.

Section~\ref{sec:model} defines the model and physical scales. Sections~\ref{sec:displacement} and \ref{sec:binned} compare the smooth displacement defect with the lattice-sensitive binned observables. Section~\ref{sec:recovery} applies the standard two-involution reduction \cite{Masanes2005,Kaniewski2016} to recover an exact Pauli algebra and derives the preparation probability and correction bounds. Sections~\ref{sec:finite} and \ref{sec:purity} treat flat and nonflat spectra; Sec.~\ref{sec:initial} compares intensity correlations. Section~\ref{sec:limit} proves the flat-spectrum oscillator limit and the positivity of its vacuum overlap. Section~\ref{sec:conclusion} discusses the conclusions and open problems, and Appendix~\ref{app:g2} corrects a published Anderson-state correlation. The hard-bin convergence rate remains open. At fixed nonzero limiting purity, convergence holds for all but countably many cutoffs; convergence and positive limiting probability at the numerical cutoffs remain unproved.

Preparation includes a joint entangling operation. A subsequent Bell violation therefore concerns the heralded prepared state, rather than the initial product vacuum. Its interpretation as a Bell test requires preparation to be completed and heralded before independently chosen local measurement settings (Sec.~\ref{sec:singlet}). The paper analyses the relation between the constituent algebra, the measurements and this specified preparation protocol; an experimental implementation remains to be developed.

\section{Paired-mode model and the Frenkel ladder}
\label{sec:model}

We consider two distinguishable fermion species and one normalised pair wavefunction of finite Schmidt rank $S$. After absorbing fixed phases into the constituent orbitals, the pair creation operator reads
\begin{equation}
B^\dagger=\sum_{k=1}^{S}\sqrt{\lambda_k}\,d_k^\dagger,\qquad d_k^\dagger=a_k^\dagger b_k^\dagger,\qquad \lambda_k>0,\quad \sum_k\lambda_k=1.
\label{eq:B}
\end{equation}
The invariant paired sector has the local basis $\ket{0}_k$, $\ket{1}_k=d_k^\dagger\ket{0}_k$. Distinct pair operators commute, $(d_k^\dagger)^2=0$ and $[d_k,d_k^\dagger]=I-2n_k$, so the sector is isomorphic to $S$ two-level systems and
\begin{equation}
[B,B^\dagger]=I-\Delta,\qquad \Delta=2\sum_k\lambda_k n_k,\qquad [\Delta,B^\dagger]=2\sum_k\lambda_k^{3/2}d_k^\dagger .
\label{eq:comm}
\end{equation}
The vacuum condition $\Delta\ket{0}=0$ does not make the Baker--Campbell--Hausdorff series of two finite rotations collapse to a phase, since $[\Delta,B^\dagger]\ket0\neq0$; this is why the Weyl relation is only approximate below. Interactions and singly occupied modes are outside the model. Alice and Bob are assigned disjoint constituent supports, so that their local algebras commute.

The unitary displacements and their Hermitian generators are
\begin{equation}
D_B(z)=e^{zB^\dagger-z^*B}=e^{iQ_z},\qquad Q_z=-i\,(zB^\dagger-z^*B),\qquad z=r,\quad w=-ib,\quad b=\frac{\pi}{2r},\quad r>0,
\label{eq:disp}
\end{equation}
i.e. $Q_r=-ir(B^\dagger-B)$ and $Q_w=-b(B^\dagger+B)$, with
\begin{equation}
[Q_r,Q_w]=-i\pi\,(I-\Delta).
\end{equation}
The canonical oscillator is obtained when the deviation operator disappears in an appropriate limit, not by setting it to zero at finite rank.

\paragraph{Flat spectrum.}
For $\lambda_k=1/S$ the states generated from the vacuum are Dicke states (the symmetric states of $S$ two-level systems with a fixed number of excitations) and the ladder closes exactly,
\begin{equation}
\begin{gathered}
\|(B^\dagger)^N\ket{0}\|^2=N!\,F_N,\qquad F_N=\frac{S!}{S^N(S-N)!},\\
B^\dagger\ket{N}=\sqrt{(N+1)\Big(1-\tfrac{N}{S}\Big)}\ket{N+1},\qquad
B\ket{N}=\sqrt{N\Big(1-\tfrac{N-1}{S}\Big)}\ket{N-1},
\end{gathered}
\label{eq:ladder}
\end{equation}
for $0\le N\le S$; these are the Frenkel-like relations of Ref.~\cite{Shiau2017} with $S=N_s$. The normalisation factor $F_N=\prod_{k=0}^{N-1}(1-k/S)$ satisfies $\log F_N=-N(N-1)/(2S)+O(N^3/S^2)$ for $N/S\to0$, reaches $F_S=S!/S^S$, and the ladder terminates at $N=S$, $B^\dagger\ket S=0$: this is the moth-eaten effect of Refs.~\cite{Pogosov2010,Shiau2017}, and it is contained in every flat-spectrum calculation below through the coefficients in \eqref{eq:ladder}. The orthogonal annihilation remainder vanishes identically. Equivalently, $B=J_-/\sqrt{S}$ for a spin $S/2$, $B+B^\dagger=2J_x/\sqrt{S}$ and $-i(B^\dagger-B)=2J_y/\sqrt{S}$: every flat-spectrum result below is a statement about functions of collective spin components of a spin $S/2$. This is the Holstein--Primakoff setting \cite{HolsteinPrimakoff1940}: the oscillator limit of Sec.~\ref{sec:limit} is the bosonic approximation of the ensemble, and the binned quadratures of Sec.~\ref{sec:binned} are sign functions of $J_x$ and $J_y$ with bin widths $\Delta m_x=r\sqrt S$ and $\Delta m_y=\pi\sqrt S/(2r)$. At $r=1$ these are respectively $2$ and $\pi$ times the projection noise $\sqrt S/2$ of the polarised state $\ket{J,-J}$, so the flat-spectrum results are exact statements about a finite spin ensemble under such a measurement \cite{Ma2011}. We keep the coboson language because it fixes the normalisation and because the nonflat case (Sec.~\ref{sec:purity}) is not a single spin. The pair-counting operator $\hat N$ differs from the field intensity,
\begin{equation}
B^\dagger B=\hat N\Big(1-\frac{\hat N-1}{S}\Big),\qquad [B,B^\dagger]=I-\frac{2\hat N}{S},
\label{eq:intensity}
\end{equation}
so the saturated state contains $S$ pairs but has intensity $\langle B^\dagger B\rangle=1$; this is the saturation side of the moth-eaten effect.

\paragraph{Physical scales.}
The Schmidt rank and purity $P=\sum_k\lambda_k^2$ (Schmidt number $\mathcal K=1/P$ \cite{Law2005}) depend on the pair wavefunction and its confinement. A Frenkel exciton delocalised uniformly over $N_s$ molecular sites gives a flat spectrum with $S=N_s$ \cite{Shiau2017}. Wannier excitons and trapped dimers -- to which the coboson formalism has been applied directly for strongly bound fermion pairs \cite{Jimenez2023} -- motivate nonflat spectra, but their Schmidt weights require a microscopic wavefunction; a finite numerical cutoff $S$ is not itself a measured coherence length or an inverse purity. Rough estimates nevertheless indicate the range: the exciton coherence length in molecular aggregates is a few to of order $10^2$ sites, and for a Wannier exciton the Schmidt number grows with the ratio of confinement size to Bohr radius (cf.\ the trapped-pair estimates of Ref.~\cite{Law2005}), from near $1$ in the strong-confinement limit, where electron and hole are confined separately and the pair wavefunction factorises, to of order $10$ in a dot a few Bohr radii across and to large values in bulk. In polariton condensates the composite nature of the exciton also becomes visible at high density, where a BCS-like description takes over from the bosonic one \cite{Goodman2023}. The calculations below therefore compare two model regimes: increasing flat rank, where the oscillator limit emerges, and increasing truncation rank at fixed nonzero limiting purity. Assigning either regime to a particular aggregate, quantum dot or atomic trap requires a system-specific Schmidt decomposition. Likewise, inferring composite quadrature statistics from emitted polariton light \cite{Deng2010} requires a coupling and detection model that we do not supply; the correlations of microcavity emission are themselves shaped by the detection optics \cite{VanRegemortel2018}.

\section{Detector (A): the response to classical drives}
\label{sec:displacement}

A classical drive displaces the composite field, and two drives with $2rb=\pi$ probe the Weyl relation. For an elementary boson $D(r)D(-ib)=-D(-ib)D(r)$; for the composite the failure of this relation on the vacuum is a function of the Schmidt spectrum that we now compute exactly. The local generators in \eqref{eq:disp} commute between different $k$, so the vacuum characteristic function factorises,
\begin{equation}
\phi_z(t)=\bra{0}e^{itQ_z}\ket{0}=\prod_k\cos\big(t|z|\sqrt{\lambda_k}\big).
\label{eq:charfun}
\end{equation}
Read as a distribution, Eq.~\eqref{eq:charfun} says that the vacuum outcomes of $Q_z$ are the signed sums $|z|\sum_k\varepsilon_k\sqrt{\lambda_k}$ with independent fair signs $\varepsilon_k$; a drive, $D_B(z)=e^{iQ_z}$, is the value of this characteristic function at $t=1$, which is the precise sense in which it probes the quadrature through its fundamental harmonic only. Define the displacement anticommutation defect for the choice $2rb=\pi$,
\begin{equation}
\delta_D=\big\|\big[D_B(r)D_B(-ib)+D_B(-ib)D_B(r)\big]\ket{0}\big\|.
\label{eq:deltaD}
\end{equation}
On a single pair mode the two displacements are the rotations
\begin{equation}
U_k=\begin{pmatrix}\cos\alpha_k&-\sin\alpha_k\\ \sin\alpha_k&\cos\alpha_k\end{pmatrix},\qquad
V_k=\begin{pmatrix}\cos\beta_k&-i\sin\beta_k\\ -i\sin\beta_k&\cos\beta_k\end{pmatrix},\qquad
\alpha_k=r\sqrt{\lambda_k},\quad\beta_k=b\sqrt{\lambda_k},
\end{equation}
with local group-commutator expectation
\begin{equation}
g_k=\bra{0}V_k^\dagger U_k^\dagger V_kU_k\ket{0}=1-2\sin^2\alpha_k\sin^2\beta_k-\frac{i}{2}\sin 2\alpha_k\,\sin 2\beta_k .
\label{eq:gk}
\end{equation}
Expanding the squared norm in \eqref{eq:deltaD} and using the product structure gives the exact expression
\begin{equation}
\delta_D^2=2+2\,\mathrm{Re}\prod_k g_k ,
\label{eq:product}
\end{equation}
without any truncation of the Baker--Campbell--Hausdorff series. Being a product of entire functions of $\sqrt{\lambda_k}$, $\delta_D^2$ is a real-analytic function of the Schmidt weights.

Let $P=\sum_k\lambda_k^2$, $M_j=\sum_k\lambda_k^j$ and $L=r^2+b^2$. On the branch of the logarithm continued from $\lambda=0$,
\begin{equation}
\begin{gathered}
\log g(\lambda)=-2irb\,\lambda+iA\lambda^2-(D+iE_3)\lambda^3+O(\lambda^4),\\
A=\tfrac43 rbL,\qquad D=2r^2b^2L,\qquad E_3=\frac{4rb(r^4+b^4)}{15}+\frac{20r^3b^3}{9}.
\end{gathered}
\label{eq:logg}
\end{equation}
Summed over modes with $\sum_k\lambda_k=1$, the first-order term gives $e^{-2irb}=-1$, the elementary-boson value of the group commutator; the defect is built entirely from the remaining orders, which is why it starts at $P^2$ and $M_3$ rather than at $P$. For fixed $r,b$ and sufficiently small largest weight $\lambda_{\max}$, analyticity bounds the summed remainder by a constant times $M_4$, and the expansion of $2-2\,\mathrm{Re}\,e^{X}$ with $X=iAP-(D+iE_3)M_3+O(M_4)$ gives
\begin{equation}
\delta_D^2=A^2P^2+2DM_3+O\big(M_4+PM_3+P^4\big).
\label{eq:expansion}
\end{equation}
The remainder is smaller than the displayed terms as $\lambda_{\max}\to0$, by $P^2\le M_3$, $M_4\le\lambda_{\max}M_3$ and $P\le\lambda_{\max}$. The third moment is therefore relevant at leading order for a general spectrum. For the flat spectrum,
\begin{equation}
\delta_D=\frac{\pi\sqrt{L+4L^2/9}}{S}+O(S^{-2}),
\label{eq:flatdefect}
\end{equation}
with coefficient $9.3252393\ldots$ at $r=1$ and minimum $8.6196987\ldots$ at the balanced scale $r=b=\sqrt{\pi/2}$. This optimisation concerns detector (A) only: at $S=8$ the balanced scale gives a vanishing preparation probability for detector (B) (Sec.~\ref{sec:finite}). The physical content of this section is that a drive, being a smooth (fundamental-harmonic) probe of the quadrature, responds to the composite through a real-analytic function of the spectrum, whose dilute expansion is governed by the first Schmidt moments; everything about detector (A) is smooth in the spectrum.

\section{Detector (B): binned homodyne and the quadrature lattice}
\label{sec:binned}

Homodyne detection of the composite field yields a quadrature outcome $q$; recording only the sign of $\cos q$ coarse-grains it into bins of width $\pi$ with alternating signs. This defines the outcome statistics of the local observables; the preparation protocols of Sec.~\ref{sec:recovery} and Appendix~\ref{app:rank4} assume in addition coherent (L\"uders) projections\footnote{The state update $\rho\mapsto P\rho P/\mathrm{Tr}(P\rho P)$ for the projector $P$ onto the selected joint eigenspace, which keeps the coherence within that eigenspace. Recording the outcome of a local homodyne measurement and binning it classically produces the same outcome statistics but not this state.} onto the joint eigenspaces of two-party observables such as $Z\otimes Z$, which ordinary homodyne detection followed by classical binning does not implement. We use the explicit bin convention
\begin{equation}
X=F(Q_r),\qquad Z=F(Q_w),\qquad F(q)=\begin{cases}+1,&\cos q\ge0,\\-1,&\cos q<0,\end{cases}\qquad
\varepsilon_v=\|(XZ+ZX)v\| .
\label{eq:bins}
\end{equation}
Both are self-adjoint involutions. For the oscillator they anticommute exactly \cite{Sorella2026}; for the coboson they do not, and $\varepsilon_v$ measures the failure on a normalised initial state $v$. The difference is the spectrum: the oscillator quadrature is continuous, whereas $Q_w$ has the $2^S$ eigenvalues $b\sum_k(-1)^{t_k}\sqrt{\lambda_k}$, counted with multiplicity -- the discrete spectrum of Fig.~\ref{fig:scheme}(b). Three structural facts follow from this and govern everything below.

\paragraph{Parity.}
$F$ is even and $Q_r,Q_w$ are odd under the pair-number parity $(-1)^{\hat N}$, so $X$ and $Z$ commute with $(-1)^{\hat N}$ and preserve number parity. For the flat spectrum the even and odd Dicke sectors can be treated separately; the quadrature $B+B^\dagger$ has eigenvalues $(S-2j)/\sqrt S$, $j=0,\dots,S$, and $X,Z$ are obtained exactly by a Krawtchouk (Dicke-basis) spectral transform. This is a reduction to an invariant physical subspace, not a numerical oscillator truncation.

\paragraph{Piecewise constancy and low moments.}
At fixed rank the spectrum of $Q_w$ consists of signed sums $bq_t$, where $q_t=\sum_k(-1)^{t_k}\sqrt{\lambda_k}$. Its product Hadamard eigenvectors do not depend on the weights; those of $Q_r$ differ by a fixed number-dependent phase. Hence $X$ and $Z$ are constant within each region of fixed bin signs, and change only at bin crossings. Expectation values and preparation probabilities inherit this property when the initial vector and other protocol parameters are fixed. A weight-dependent initial state need not do so.

Purity and the third moment do not determine the binned observables. The rank-four spectra
\begin{equation}
\lambda^{(a)}=(0.49,\,0.01,\,0.25,\,0.25),\qquad
\lambda^{(b)}=(0.442,\,0.058,\,0.394,\,0.106)
\label{eq:pair}
\end{equation}
have $P=913/2500$ and $M_3=1489/10000$, but at $r=1$ their squared vacuum defects are $1$ and $5/2$, respectively. The difference is visible in the bins: for $\lambda^{(a)}$ the two drives cut the set of signed sums at the same places -- the signs $F(rq_t)$ and $F(bq_t)$ coincide for every $t$, both being $-1$ exactly at $|q_t|=1.6$ and $1.8$ -- whereas for $\lambda^{(b)}$ they cut it at different ones ($F(rq_t)=-1$ only at $|q_t|=1.86$, $F(bq_t)=-1$ at $|q_t|=1.21$, $1.38$ and $1.86$). A specified sequential preparation also gives different probabilities and Bell values (Appendix~\ref{app:rank4}). These spectra are outside the dilute regime of \eqref{eq:expansion}; they establish nondetermination by these two moments, not a failure of that expansion. They do not show that moments cannot give bounds. Indeed, at known finite rank, the power sums through order $S$ reconstruct the unordered spectrum by Newton's identities, and determine the vacuum protocol up to mode relabeling.

\paragraph{High harmonics and edge projectors.}
Away from its discontinuities the square wave has the Fourier series $F(q)=\frac4\pi\sum_{m\ge0}\frac{(-1)^m}{2m+1}\cos(2m+1)q$. At each bin edge $c\in\mathcal C=\pi/2+\pi\mathbb Z$, every term vanishes, whereas our convention assigns $F(c)=+1$. If $F_M^{\rm F}$ denotes the Fej\'er approximation\footnote{The Ces\`aro means of the Fourier partial sums. Unlike the partial sums, which overshoot near a jump (Gibbs phenomenon), the Fej\'er sums of a function bounded by $1$ are themselves bounded by $1$, which is what allows the limit to be taken operator by operator in Sec.~\ref{sec:limit}.}, the spectral theorem gives, with $\operatorname{s-lim}$ denoting the strong operator limit (convergence on every vector),
\begin{equation}
F(Q)=\operatorname*{s-lim}_{M\to\infty}F_M^{\rm F}(Q)+P_{\rm edge}(Q),\qquad
P_{\rm edge}(Q)=\mathbf1_{\mathcal C}(Q).
\label{eq:edgeFourier}
\end{equation}
The last term has a minus sign for the alternative edge convention. It vanishes for the oscillator spectral measure, but generally not at finite rank. With $X=\bar X+P_r$, $Z=\bar Z+P_w$, where the bars denote the Fej\'er limits, the anticommutator includes
\begin{equation}
\{X,Z\}=\{\bar X,\bar Z\}+\{\bar X,P_w\}+\{P_r,\bar Z\}+\{P_r,P_w\}.
\label{eq:edgeAnti}
\end{equation}
A Fourier-only estimate cannot omit these terms at a finite-rank edge crossing. Even without edge atoms, low-order displacement estimates are not uniform over all harmonics. Multiplying the limiting Gaussian by a position square wave also creates jumps and takes the vector outside the kinetic-energy form domain, so finite-occupation error estimates cannot be propagated through the measurement sequence without further control.

\subsection{Stability and a smooth measurement model}
\label{sec:robustness}
Let $\gamma_a=\min_t\operatorname{dist}(a q_t,\mathcal C)$ for $a=r,b$. These margins belong to the finite spectrum; $0\le\gamma_a\le\pi/2$.
\begin{proposition}
\label{prop:margin}
Consider two spectra of the same rank in the same paired basis, with scales $(r,b)$ and $(r',b')$. Put
\begin{equation}
d=\sum_k|\sqrt{\lambda'_k}-\sqrt{\lambda_k}|,\qquad
\zeta_a=a'd+|a'-a|\sum_k\sqrt{\lambda_k},\quad a\in\{r,b\}.
\label{eq:margin}
\end{equation}
If $\zeta_r<\gamma_r$ and $\zeta_b<\gamma_b$, both binned operators are unchanged. Thus their recovery protocol is unchanged for a fixed initial vector and cutoff.
\end{proposition}
\begin{proof}
For every sign string, $|a'q'_t-aq_t|\le\zeta_a$ by the triangle inequality. Under the strict margin condition no eigenvalue crosses a bin edge. The product eigenvectors are fixed, so identical signs give identical operators.
\end{proof}

For a specified unsharp version, take $F_\eta(q)=\tanh(\cos q/\eta)$, $\eta>0$, and write $X_\eta=F_\eta(Q_r)$, $Z_\eta=F_\eta(Q_w)$. These Hermitian contractions define binary POVMs $(I\pm X_\eta)/2$ and $(I\pm Z_\eta)/2$. With
\begin{equation}
h_a=1-\tanh\!\left(\frac{\sin\gamma_a}{\eta}\right),\qquad a=r,b,
\label{eq:smoothh}
\end{equation}
functional calculus gives
\begin{equation}
\begin{gathered}
\|X_\eta-X\|\le h_r,\qquad \|Z_\eta-Z\|\le h_b,\\
\|\{X_\eta,Z_\eta\}-\{X,Z\}\|\le2(h_r+h_b),\\
\|I-X_\eta^2\|\le2h_r-h_r^2,\qquad
\|I-Z_\eta^2\|\le2h_b-h_b^2.
\end{gathered}
\label{eq:smoothbounds}
\end{equation}
Indeed, $|\cos(aq_t)|\ge\sin\gamma_a$; subtracting the products $X_\eta Z_\eta-XZ$ and $Z_\eta X_\eta-ZX$, and using the contraction bounds, proves the second line. For positive margins, $h_a\le2e^{-2\sin\gamma_a/\eta}$, whereas at an edge $F_\eta=0$ and its difference from the hard convention is one. The bound consequently does not become small there. A small smoothed anticommutator alone is not a Pauli algebra: the unsharpness bounds must also be considered. The exact recovery below applies to the hard involutions, not unchanged to these POVMs.

\section{Exact Pauli recovery and its efficiency}
\label{sec:recovery}

\subsection{Reducing spectral filter and polar correction}

Let $X,Z$ be self-adjoint involutions on a Hilbert space and set
\begin{equation}
K=\frac{XZ+ZX}{2},\qquad R=Z-KX,\qquad E=\mathbf 1_{[-\kappa,\kappa]}(K),\qquad 0<\kappa<1 .
\label{eq:KRE}
\end{equation}
$K$ commutes with $X$ and $Z$; direct multiplication gives $\{X,R\}=0$ and $R^2=I-K^2$. The range of $E$ reduces $X$, $Z$ and $K$, and on it $(I-K^2)^{-1/2}$ is bounded. Define there
\begin{equation}
\Zt=R\,(I-K^2)^{-1/2},\qquad \Yt=iX\Zt .
\label{eq:polar}
\end{equation}
These are self-adjoint involutions with $\{X,\Zt\}=0$, hence $(X,\Yt,\Zt)$ is an exact Pauli algebra on $\mathrm{ran}\,E$, generally with multiplicity. This is the Jordan two-involution reduction (Halmos's two-subspace theorem \cite{Halmos1969}). In finite dimensions, $X$ and $Z$ decompose into two-dimensional blocks on which they act as $\sigma_x$ and $\cos\theta\,\sigma_x+\sin\theta\,\sigma_z$, with $K=\cos\theta$; the filter removes the one-dimensional blocks with $K=\pm1$. In general, the spectral decomposition may be a direct integral, with multiplicity, rather than a discrete sum. The bounded operator identities above apply in either case and give the representation explicitly: since $X$ and $\Zt$ are anticommuting involutions, $\mathrm{ran}\,E\cong\mathbb C^2\otimes\mathcal M$ with $X=\sigma_x\otimes I$ and $\Zt=\sigma_z\otimes I$; since $K$ commutes with both, $K=I\otimes C$ for a self-adjoint $C$ on $\mathcal M$ with $\|C\|\le\kappa$; and then $Z=KX+R=\sigma_x\otimes C+\sigma_z\otimes\sqrt{I-C^2}$. The factor $\mathcal M$ is the complementary logical factor referred to in the Introduction. We claim no novelty for this algebraic step, which is the standard tool of self-testing \cite{Masanes2005,Kaniewski2016,Supic2020}; the content here is its application to specified coboson quadratures and the calculation of the rejected weight, the measurement correction and the success probability.

The rejected weight and the correction are controlled by the original defect. For a normalised initial state $v$ let $t=\bra vE\ket v$. Since $\varepsilon_v=2\|Kv\|$, Chebyshev's inequality gives
\begin{equation}
t\ge\max\Big(0,\,1-\frac{\varepsilon_v^2}{4\kappa^2}\Big).
\label{eq:tbound}
\end{equation}
Moreover, on the retained sector $\{\Zt,Z\}=2\sqrt{I-K^2}$, so
\begin{equation}
(\Zt-Z)^2=2\big(I-\sqrt{I-K^2}\big),\qquad
\|\Zt-Z\|_{E}\le\sqrt{2\big(1-\sqrt{1-\kappa^2}\big)},\qquad
\|(\Zt-Z)Ev\|^2\le\varepsilon_v^2/2,
\label{eq:corrbound}
\end{equation}
the last using $1-\sqrt{1-k^2}\le k^2$. A smaller cutoff tightens the operator correction but may reject more weight. It should be kept in mind that $\Zt$ is a nonlinear function of \emph{both} conjugate quadratures; at finite rank it is not a binned quadrature measurement, and realising it requires coherent control of a two-quadrature observable.

\subsection{Singlet projector, success probability, and a mixedness criterion}
\label{sec:singlet}

Extend the corrected operators by zero outside the retained sector and write $X_E=EXE$. The joint observables $X_E\otimes X_E$ and $\Zt\otimes\Zt$ commute, and their simultaneous minus-sector projector is
\begin{equation}
P_-^{(\kappa)}=\frac14\big(E\otimes E-X_E\otimes X_E-\Yt\otimes\Yt-\Zt\otimes\Zt\big).
\label{eq:Pminus}
\end{equation}
For identical normalised local initial states $v$, with $m_i=\bra v\sigma_{i,E}\ket v$, $\sigma_{i,E}\in\{X_E,\Yt,\Zt\}$, idempotence gives the total probability including both local filters,
\begin{equation}
p(v)=\big\|P_-^{(\kappa)}(v\otimes v)\big\|^2=\frac14\big(t^2-m_x^2-m_y^2-m_z^2\big).
\label{eq:pv}
\end{equation}
For $p>0$ every normalised projected state has $\langle\sigma_{i,A}\sigma_{j,B}\rangle=-\delta_{ij}$ and vanishing local Pauli means; with Alice's axes $X_E,\Zt$ and Bob's $(X_E\pm\Zt)/\sqrt2$ the CHSH value \cite{CHSH1969} is $2\sqrt2$, Tsirelson's bound \cite{Cirelson1980}, as for any exact Pauli pair in a singlet. Within the theory this value therefore carries no information about the composite; the model-dependent quantities are $t$, the correction \eqref{eq:corrbound}, and $p$. Its role is experimental: by the self-testing theorems \cite{Kaniewski2016,Supic2020}, an observed CHSH value of $2\sqrt2$ certifies, without any model of the devices or of the composite, that the prepared state is a singlet and the measurements are a Pauli pair on its support, up to local isometries\footnote{That is, there exist isometries acting on Alice's and Bob's spaces separately that map the prepared state and the measured observables onto a singlet and a Pauli pair, tensored with an uncorrelated remainder. Sec.~\ref{sec:finite} gives the quantitative version (extractability).}; a Bell test therefore verifies the output of an implementation of the filter and polar correction, though not the internal operations that produced it. The Bell value is informative within the theory only for the \emph{uncorrected} protocol, whose conditional state is not a singlet: there it measures, in the units of a Bell experiment, how far the binned quadratures are from a Pauli pair, and the same self-testing bounds turn it into a certified singlet fidelity (Sec.~\ref{sec:finite}, Appendix~\ref{app:rank4}).

For $t>0$, decompose $\mathrm{ran}\,E\cong\mathbb C^2\otimes\mathcal M$ into qubit and multiplicity factors. The filtered pure state $Ev/\sqrt t$ has reduced qubit state $\rho_v$ with Bloch vector $\vec m/t$, and
\begin{equation}
p(v)=\frac{t^2}{4}\left(1-|\vec m/t|^2\right)
=\frac{t^2}{2}\left(1-\Tr\rho_v^2\right).
\label{eq:mixed}
\end{equation}
This is the two-copy identity for the antisymmetric qubit projector $(I-\mathrm{Swap})/2$, used in purity and concurrence measurements \cite{Walborn2007}. Here it separates the accepted-weight factor $t^2$ from the conditional success probability. For $t=0$, $p=0$ and the normalized filtered state is undefined. For $t>0$, preparation succeeds if and only if $\rho_v$ is mixed, or equivalently the filtered pure state is entangled between its qubit and multiplicity factors. A state in one Jordan block cannot succeed. Merely occupying several blocks is insufficient: $\ket\chi\otimes(\ket a+\ket b)/\sqrt2$ still has a pure reduced qubit. The rank-four spectrum $\lambda^{(a)}$ of \eqref{eq:pair} is an instance: at $r=1$ its $K$ equals $-1/2$ on the whole vacuum, so $t=1$ at $\kappa=1/2$, yet $p=0$, because the vacuum is a product of a qubit state and a multiplicity state across four blocks of equal angle.

The factor $\mathcal M$ is a logical subsystem defined by the two observables -- the label of the Jordan block -- not a spatial or constituent subsystem; the statement that the filtered vacuum is entangled between the qubit and $\mathcal M$ is a statement in this factorisation, and the same structure exists for the elementary oscillator, whose vacuum at $r=1$ has a reduced-qubit Bloch length $0.85$. What is specific to the composite is how the block weights, and with them $t$ and $p$, depend on the Schmidt spectrum. A composite whose filtered vacuum sits in a single block, however large its accepted weight, yields nothing. The initial-state spectral measure of $K$ determines $t$, but not in general the three Pauli means or $p$. For instance, $X=\sigma_x\otimes I$, $Z=\sigma_z\otimes I$ give $K=0$ and $t=1$ for every initial state. A product state between qubit and multiplicity has $p=0$, whereas a maximally entangled state of those factors has $p=1/4$, with the same $K$ spectral measure.

\begin{lemma}
\label{lem:cutoff}
For fixed $X,Z,v$, the total probability $p_\kappa(v)$ is nondecreasing in $0<\kappa<1$.
\end{lemma}
\begin{proof}
On $E_0=\mathbf1_{(-1,1)}(K)$ the polar part of $R$ defines a bounded corrected Pauli representation. Each $E_\kappa$ reduces it, and the corresponding singlet projector is the restriction of the same projector $P_-^0$:
$P_-^{(\kappa)}=(E_\kappa\otimes E_\kappa)P_-^0$. These commuting projections are nested as $\kappa$ increases. Their expectation in $v\otimes v$ is therefore nondecreasing.
\end{proof}
The jump locations at finite rank come from $|K|$, but their heights depend on the initial qubit structure as well as its spectral weights. Neither a nonzero reducing sector nor positive $t$ alone guarantees $p>0$. Conditional CHSH saturation assumes heralded preparation before the measurement settings are chosen; for massive-particle realizations, number-changing operations would additionally require an appropriate phase reference or reservoir.

\section{Flat spectrum: how fast the two detectors stop seeing the composite}
\label{sec:finite}

\begin{figure}[t]
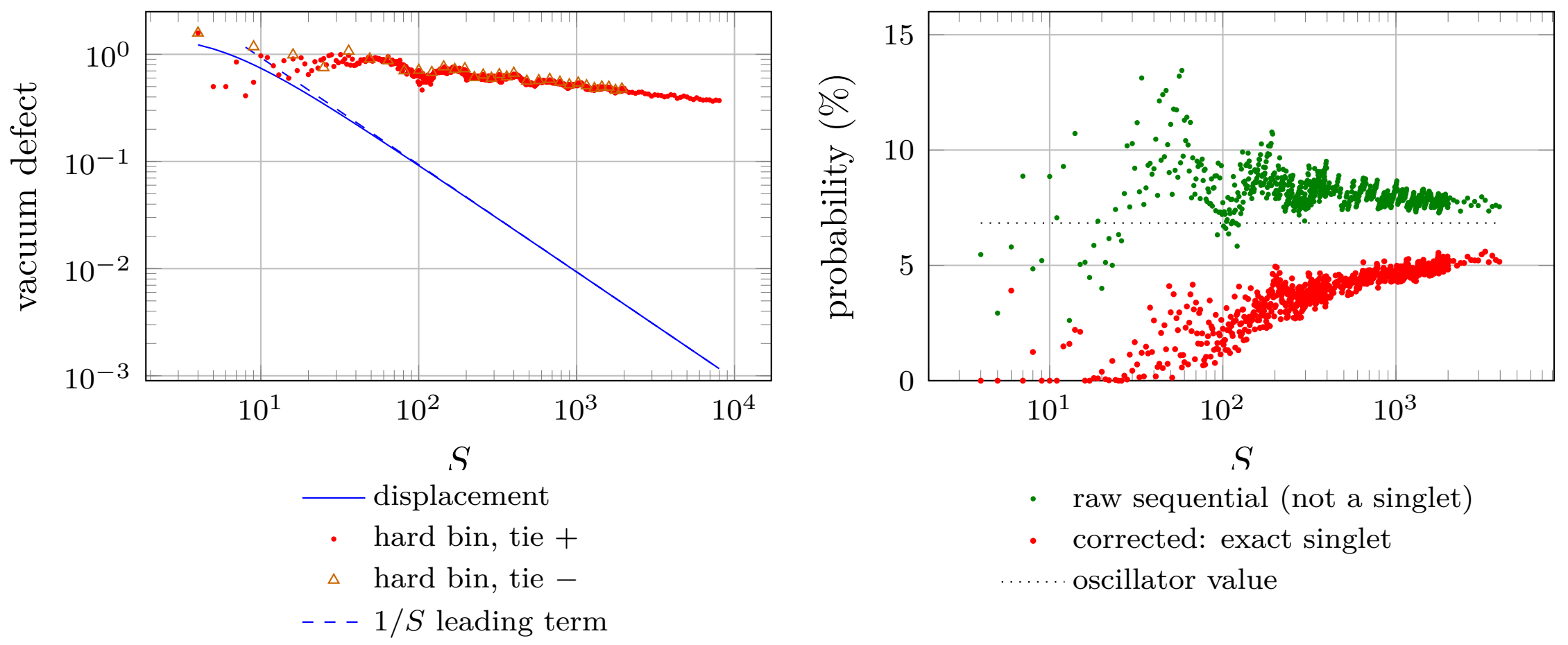

\centering

\caption{The two detectors on the Frenkel ladder, vacuum initial state, $r=1$: every integer rank $4\le S\le400$, every fourth rank to $2000$, and $30$ log-spaced ranks to $8000$. (a) Vacuum anticommutation defects. The drive (blue, $\delta_D$) follows the $1/S$ law \eqref{eq:flatdefect}: it sees the composite through the commutator deviation $\Delta=2\hat N/S$, which tends to zero at fixed pair number as the ladder opens. Binned homodyne (red, $\varepsilon_{\rm vac}$) sees the lattice of Fig.~\ref{fig:scheme}(b), whose spacing $\pi/\sqrt S$ shrinks slowly against the fixed bin width; its defect is still $0.37$ at $S=8000$, scatters about its trend by up to $\pm15\%$ as lattice points cross bin edges, and at perfect squares (triangles: $-1$ convention; dots: $+1$) depends on the sign assigned to the points that sit exactly on an edge. (b) Probability of projecting two vacua into the singlet sector, $+1$ convention. Red: the corrected protocol of Sec.~\ref{sec:recovery} ($\kappa=1/2$, to $S=4096$), which prepares an exact singlet; it vanishes at the ranks where the filtered vacuum has a pure reduced qubit and approaches the oscillator value $p_\infty(1)=0.0683$ (dotted) as slowly as the red curve in (a). Green: the raw sequential protocol of Appendix~\ref{app:rank4}, the probability of obtaining the two ``$-$'' outcomes in succession; it is larger, often above the oscillator value, but since $X$ and $Z$ do not anticommute the conditional state is not a singlet, so this is not a success probability. All points are exact finite-rank values up to floating-point error.}
\label{fig:scan}
\end{figure}

For a Frenkel-like coboson the number of sites $S$ is the only parameter, and both detectors (A) and (B) must eventually see an elementary boson, since the ladder \eqref{eq:ladder} tends to the oscillator one. This section follows them up to $S=8000$. The result, Fig.~\ref{fig:scan}, is that the drive stops seeing the composite as $1/S$, while binned homodyne still sees it at $S=8000$.

Figure~\ref{fig:scan} and Table~\ref{tab:scan} give the vacuum calculations at the natural scale $r=1$ and cutoff $\kappa=1/2$, for all ranks up to $400$ and a selection up to $8000$. Panel (a) should be read as a comparison of resolutions: both curves start at order one at $S\sim10$, where the lattice has a handful of points per bin, but at $S=1000$ the blue curve lies a factor of about fifty below the red one ($0.0093$ against $0.52$). Panel (b) shows what this means for the task: the singlet probability of the corrected protocol follows the slow curve, not the fast one. The features are the following.

\begin{table}[t]
\centering
\footnotesize
\setlength{\tabcolsep}{4pt}
\begin{tabular}{lrrrrrrrrr}
\toprule
$S$ & tie & $\delta_D$ & $\varepsilon_0$ & $t$ & $|\vec m|/t$ & $p_{\rm corr}$ & $p_{\rm raw}$ & $\mathcal B_{\rm raw}$ & edges\\
\midrule
8 & $+$ & 0.8652 & 0.4104 & 0.9698 & 0.9732 & 0.012425 & 0.0485 & 2.360 & --\\
12 & $+$ & 0.6458 & 0.7825 & 0.9531 & 0.9667 & 0.014866 & 0.0929 & 2.443 & --\\
16 & $+$ & 0.5109 & 0.9059 & 0.8436 & 0.9999 & 0.000024 & 0.0513 & 2.379 & 4\\
16 & $-$ & 0.5109 & 0.9881 & 0.3988 & 1.0000 & 0.000000 & 0.0924 & 2.417 & 4\\
24 & $+$ & 0.3579 & 0.8839 & 0.7756 & 0.9990 & 0.000310 & 0.0742 & 2.377 & --\\
32 & $+$ & 0.2747 & 0.9933 & 0.7433 & 0.9742 & 0.007038 & 0.1119 & 2.287 & --\\
64 & $+$ & 0.1418 & 0.9506 & 0.7504 & 0.9063 & 0.025149 & 0.0909 & 2.322 & 8\\
64 & $-$ & 0.1418 & 0.8705 & 0.8122 & 0.9019 & 0.030765 & 0.1209 & 2.296 & 8\\
128 & $+$ & 0.0719 & 0.6750 & 0.8392 & 0.9267 & 0.024860 & 0.0918 & 2.465 & --\\
256 & $+$ & 0.0362 & 0.6472 & 0.8670 & 0.9110 & 0.031950 & 0.0767 & 2.511 & 16\\
256 & $-$ & 0.0362 & 0.6409 & 0.8652 & 0.9000 & 0.035554 & 0.0935 & 2.462 & 16\\
512 & $+$ & 0.0182 & 0.5196 & 0.9047 & 0.8853 & 0.044264 & 0.0767 & 2.583 & --\\
1024 & $+$ & 0.0091 & 0.5404 & 0.8982 & 0.8838 & 0.044143 & 0.0765 & 2.572 & 32\\
1024 & $-$ & 0.0091 & 0.5397 & 0.8927 & 0.8854 & 0.043058 & 0.0848 & 2.543 & 32\\
2050 & $+$ & 0.0045 & 0.4574 & 0.9350 & 0.8732 & 0.051896 & 0.0793 & 2.623 & --\\
3956 & $+$ & 0.0024 & 0.4191 & 0.9422 & 0.8762 & 0.051563 & 0.0754 & 2.648 & --\\
8000 & $+$ & 0.0012 & 0.3707 & -- & -- & -- & 0.0753 & 2.680 & --\\
$\infty$ & -- & 0 & 0 & 1 & 0.8522 & 0.068350 & 0.0683 & 2.828 & --\\
\bottomrule
\end{tabular}
\caption{Flat spectrum, vacuum initial state, $r=1$, $\kappa=1/2$. For every perfect-square rank shown the paired entries are for the tie-breaking conventions $F=+1$ / $F=-1$ at exact bin-edge eigenvalues; ``edge atoms'' is the number of $Q_w$ eigenvalues lying exactly on a bin edge. $\mathcal B_{\rm raw}$ is the Bob-optimised CHSH value of the raw sequential state (Appendix~\ref{app:rank4}); the corrected protocol gives $2\sqrt2=2.828$ at every rank with $p_{\rm corr}>0$. $|\vec m|/t$ is the Bloch length of the reduced qubit state of the filtered vacuum, Eq.~\eqref{eq:mixed}. The last row is the oscillator limit of Sec.~\ref{sec:limit}.}
\label{tab:scan}
\end{table}

\paragraph{The two defects converge at different observed rates.}
The displacement defect follows the controlled asymptotic law \eqref{eq:flatdefect}; its leading term is 6.6\% above the exact result at $S=30$, 2.8\% at $64$, and 0.65\% at $256$. The hard-bin defect remains of order one at many ranks near $100$ and is approximately $0.37$ at $S=8000$, where $\delta_D\simeq0.0012$. Unweighted log--log fits over the scan ($+1$ edge convention) give slopes $-0.146$ for $64\le S\le8000$, $-0.154$ for $256\le S\le8000$, and $-0.163$ for $1024\le S\le8000$; these are finite-window slopes. The scatter about the fitted trend is about $\pm15\%$ for $64\le S\le400$ (neighbouring ranks differ by up to $40\%$ near $S=100$), $\pm6\%$ for the sampled ranks between $400$ and $2000$, and $\pm5\%$ above; it narrows slowly while the trend stays shallow.

\paragraph{Harmonic sensitivity without a rate claim.}
For odd harmonic orders $u=2m+1$, $v=2n+1$, the expansion of the single-mode rotations suggests the leading phase-error scale
\begin{equation}
O\!\left(\frac{uv(u^2r^2+v^2b^2)}{S}\right).
\label{eq:heur}
\end{equation}
Comparable orders give a crossover scale $S^{1/4}$, while fixing one order gives $S^{1/3}$ for the other, with scale-dependent constants. Neither observation yields a bound on the full square-wave anticommutator, because cross terms, the nonuniform vacuum spectral measure, aliasing, and the edge projectors in \eqref{eq:edgeAnti} would all have to be controlled; in particular the weight of the square wave on harmonics beyond the crossover cannot simply be identified with the vacuum defect, so we do not turn the crossover scale into a rate. Directly computed harmonic defects are given in Appendix~\ref{app:harmonics}; they show that the fundamental-harmonic (displacement) approximation is not uniform in the harmonic order, which is the mechanism behind the slow convergence.

\paragraph{Exact edge atoms at perfect squares.}
At $r=1$ the $Q_w$ eigenvalue $b(S-2j)/\sqrt S$ equals a bin edge $(2m+1)\pi/2$ whenever $(S-2j)/\sqrt S=2m+1$. If $S=n^2$ this has the integer solutions $j=(n^2-(2m+1)n)/2$, $2m+1\le n$, so every perfect-square rank has spectral atoms exactly at bin edges ($2\lceil n/2\rceil$ of them: four at $S=9,16$, six at $S=25,36$, eight at $S=49,64$), while $Q_r$ never does at $r=1$ because $\pi/2$ is irrational. At these ranks the convention $F(\pm\pi/2)=+1$ in \eqref{eq:bins} does real work: at $S=16$ switching from the $+1$ to the $-1$ convention changes $t$ from $0.84$ to $0.40$ and $p$ from $2.4\times10^{-5}$ to $0$; at $S=9$ the alternative convention gives $t=0$. For other scales $r$ the same phenomenon occurs at other ranks. Any statement about a specific finite rank must therefore name the convention, and the perfect-square points in Fig.~\ref{fig:scan} should be read as convention dependent.

\paragraph{Preparation probability.}
$p_{\rm corr}$ vanishes to machine precision at the ranks $S=4$, $5$, $7$, $9$, $10$, $11$, $17$ and $25$, where the Bloch length in Table~\ref{tab:scan} equals $1$: the filtered vacuum has a pure reduced qubit and the criterion \eqref{eq:mixed} fails. It is $2.4\times10^{-5}$ at $S=16$ and $2.5\times10^{-6}$ at $S=26$, $1.2\%$ at $S=8$ (the exact construction of Appendix~\ref{app:eight}; a favourable small rank, not a representative one), below $0.1\%$ for most of $16\le S\le28$, between $0.1\%$ and $4.2\%$ for $32\le S\le127$, between $1.7\%$ and $4.9\%$ for $128\le S\le255$, and between $4.3\%$ and $5.6\%$ for the sampled ranks $S>1024$; near $S=4000$ it is $5.2\%$, still $25\%$ below the oscillator limit $6.8\%$. The raw sequential preparation has a different finite-rank success probability and does not prepare an exact singlet; both protocols approach the same oscillator limit.

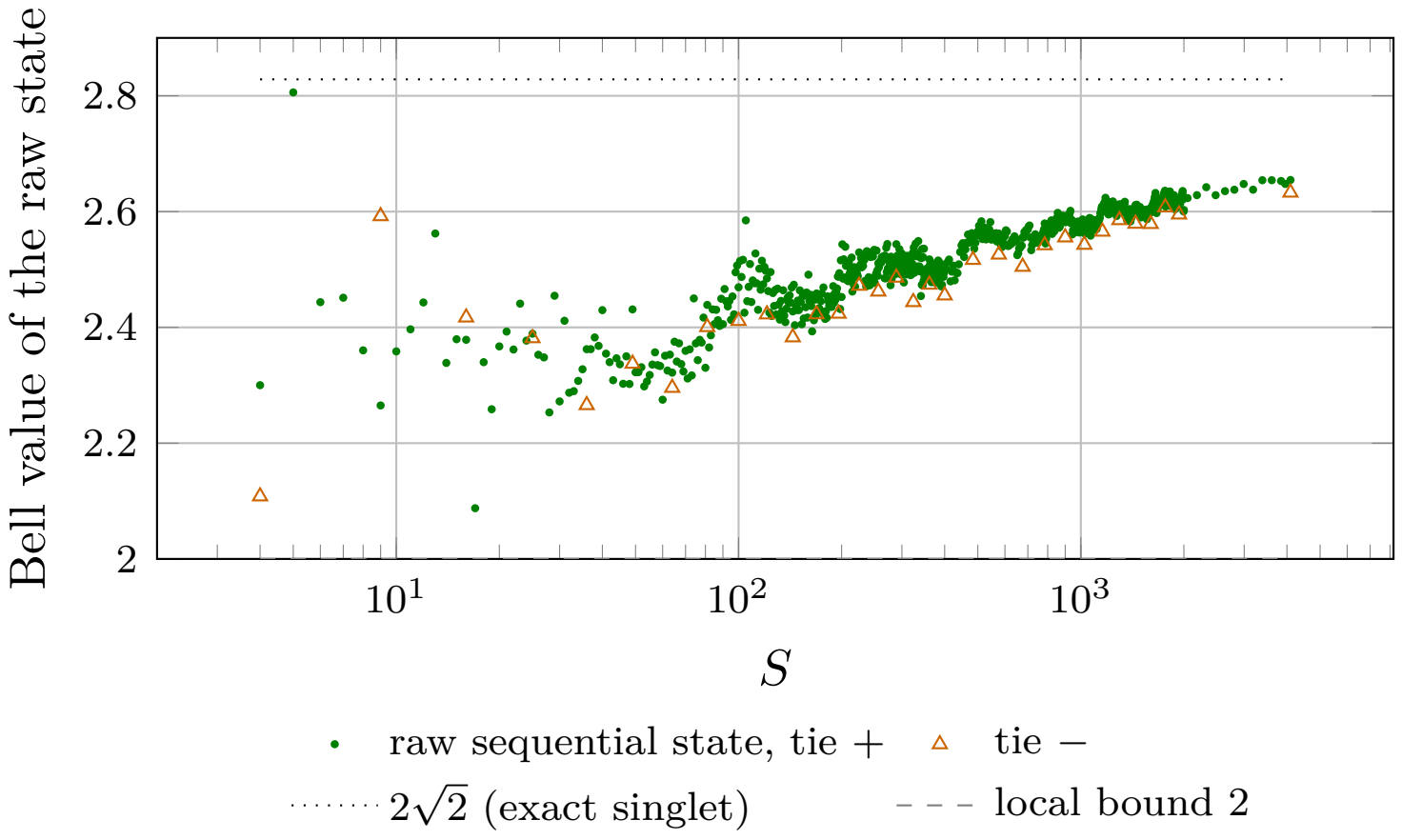
\begin{figure}[t]
\centering
\begin{tikzpicture}
\begin{axis}[width=0.62\textwidth,height=5.2cm,xmode=log,xlabel={$S$},ylabel={Bell value of the raw state},ymin=2,ymax=2.9,grid=major,tick label style={font=\scriptsize},label style={font=\small},legend style={font=\scriptsize,at={(0.5,-0.28)},anchor=north,draw=none,legend columns=2},legend cell align=left]
\addplot[green!50!black,only marks,mark=*,mark size=.6pt] coordinates {(4,2.30015064) (5,2.8058837) (6,2.44356887) (7,2.45121353) (8,2.36030946) (9,2.2651549) (10,2.35853068) (11,2.396649) (12,2.4430608) (13,2.56209552) (14,2.33861304) (15,2.37948364) (16,2.37856272) (17,2.08765188) (18,2.33987112) (19,2.25862168) (20,2.36715868) (21,2.39272867) (22,2.36174439) (23,2.44100361) (24,2.37710631) (25,2.38903635) (26,2.35269187) (27,2.34811793) (28,2.25308043) (29,2.45462951) (30,2.27210827) (31,2.41143694) (32,2.2870982) (33,2.2898583) (34,2.30757136) (35,2.32764666) (36,2.36231194) (37,2.36239584) (38,2.38262462) (39,2.36805506) (40,2.42974386) (41,2.35456794) (42,2.33991342) (43,2.30872077) (44,2.34657513) (45,2.3362481) (46,2.30256765) (47,2.3500028) (48,2.30225071) (49,2.43105362) (50,2.32246224) (51,2.32270311) (52,2.33115138) (53,2.29802453) (54,2.30647547) (55,2.31775623) (56,2.3357242) (57,2.35689597) (58,2.33479626) (59,2.3332876) (60,2.27508439) (61,2.35100537) (62,2.32544466) (63,2.35291375) (64,2.32183258) (65,2.37524509) (66,2.34103947) (67,2.37242733) (68,2.33662156) (69,2.32385448) (70,2.35942313) (71,2.3117341) (72,2.36211986) (73,2.31696866) (74,2.45004312) (75,2.37232577) (76,2.343392) (77,2.37848641) (78,2.37382013) (79,2.416907) (80,2.33022462) (81,2.43882097) (82,2.36531391) (83,2.38634086) (84,2.43105311) (85,2.40559312) (86,2.43031158) (87,2.41241703) (88,2.40251757) (89,2.44951149) (90,2.4055323) (91,2.466273) (92,2.4450467) (93,2.4362123) (94,2.44563978) (95,2.41297698) (96,2.45343148) (97,2.42253488) (98,2.49515217) (99,2.50691447) (100,2.46927544) (101,2.51469023) (102,2.48708953) (103,2.51674417) (104,2.42532476) (105,2.58481623) (106,2.44528804) (107,2.47003851) (108,2.50944814) (109,2.44381801) (110,2.4811167) (111,2.47636184) (112,2.52761881) (113,2.47612389) (114,2.43034052) (115,2.50139287) (116,2.46512228) (117,2.51508661) (118,2.47434653) (119,2.50445064) (120,2.49786565) (121,2.48450361) (122,2.46274074) (123,2.42550622) (124,2.49552633) (125,2.41972419) (126,2.4593125) (127,2.43697689) (128,2.46460847) (129,2.42237643) (130,2.46689507) (131,2.43942975) (132,2.41337851) (133,2.44574203) (134,2.42782244) (135,2.46405273) (136,2.42051694) (137,2.4090953) (138,2.43239233) (139,2.4512311) (140,2.44351006) (141,2.45532938) (142,2.44279209) (143,2.46888001) (144,2.4275809) (145,2.47362036) (146,2.40375603) (147,2.43754896) (148,2.42996753) (149,2.44631526) (150,2.42990219) (151,2.46389234) (152,2.44351103) (153,2.40541993) (154,2.45377844) (155,2.41538278) (156,2.45561358) (157,2.4348688) (158,2.46023381) (159,2.41701533) (160,2.49102895) (161,2.44595008) (162,2.45633655) (163,2.43688971) (164,2.39311793) (165,2.45545288) (166,2.41232108) (167,2.44781182) (168,2.42074618) (169,2.4585781) (170,2.42561794) (171,2.42558055) (172,2.43293945) (173,2.45633326) (174,2.44517677) (175,2.45871511) (176,2.42714975) (177,2.46896441) (178,2.42042877) (179,2.41455437) (180,2.42809089) (181,2.4166058) (182,2.45364985) (183,2.44546327) (184,2.43750832) (185,2.4517931) (186,2.43972615) (187,2.43533149) (188,2.45400874) (189,2.47118567) (190,2.44240693) (191,2.4586219) (192,2.44796896) (193,2.46024952) (194,2.46716365) (195,2.48791138) (196,2.48768005) (197,2.49357033) (198,2.43176082) (199,2.51585383) (200,2.45248226) (201,2.54345635) (202,2.48414807) (203,2.51763656) (204,2.48624317) (205,2.53921964) (206,2.50633584) (207,2.48192317) (208,2.51104879) (209,2.47982116) (210,2.51167309) (211,2.49296121) (212,2.50199356) (213,2.50240671) (214,2.5205957) (215,2.46194917) (216,2.48513861) (217,2.47334232) (218,2.50395663) (219,2.47033661) (220,2.49752397) (221,2.48815521) (222,2.49225734) (223,2.488239) (224,2.50341596) (225,2.52944584) (226,2.50479046) (227,2.50838742) (228,2.5195023) (229,2.50378015) (230,2.51242493) (231,2.50614248) (232,2.51961423) (233,2.51656302) (234,2.48037018) (235,2.52277516) (236,2.47093184) (237,2.5315996) (238,2.47270355) (239,2.5238544) (240,2.48660918) (241,2.52694771) (242,2.47280608) (243,2.5215254) (244,2.48476679) (245,2.52991608) (246,2.48459431) (247,2.52241162) (248,2.48199441) (249,2.53387307) (250,2.47670529) (251,2.53274039) (252,2.49230495) (253,2.53260172) (254,2.49776324) (255,2.48909366) (256,2.51074842) (257,2.50356133) (258,2.50925253) (259,2.51248246) (260,2.51515334) (261,2.51176693) (262,2.52130137) (263,2.51671529) (264,2.51923229) (265,2.51933228) (266,2.52344145) (267,2.48489772) (268,2.52384727) (269,2.48470433) (270,2.53763111) (271,2.49571414) (272,2.54391799) (273,2.50181604) (274,2.50032949) (275,2.51385239) (276,2.51443469) (277,2.51093896) (278,2.47284173) (279,2.52210224) (280,2.48393865) (281,2.51223508) (282,2.49501554) (283,2.50176654) (284,2.49745491) (285,2.5131019) (286,2.4916577) (287,2.51361397) (288,2.50978977) (289,2.54362922) (290,2.49310903) (291,2.52399406) (292,2.50517691) (293,2.52227262) (294,2.50501973) (295,2.53509528) (296,2.51650236) (297,2.50173815) (298,2.51591198) (299,2.49114455) (300,2.52740983) (301,2.51110703) (302,2.50020604) (303,2.50676547) (304,2.47757508) (305,2.50746354) (306,2.49108553) (307,2.51923227) (308,2.49566141) (309,2.5216791) (310,2.49165372) (311,2.52368953) (312,2.51326104) (313,2.49914815) (314,2.50471864) (315,2.51370591) (316,2.51979649) (317,2.51426716) (318,2.4902118) (319,2.5111946) (320,2.49150975) (321,2.52097786) (322,2.50983413) (323,2.52884796) (324,2.49558897) (325,2.53807201) (326,2.49418518) (327,2.53024858) (328,2.50382742) (329,2.54469822) (330,2.50329176) (331,2.52842786) (332,2.52288399) (333,2.52776753) (334,2.50534088) (335,2.54908698) (336,2.52294404) (337,2.49763905) (338,2.50506728) (339,2.49675505) (340,2.52058946) (341,2.45415031) (342,2.51955663) (343,2.47360038) (344,2.53537374) (345,2.4724371) (346,2.52123858) (347,2.48006767) (348,2.53593305) (349,2.48872693) (350,2.48483995) (351,2.48710585) (352,2.49886761) (353,2.49049528) (354,2.50104596) (355,2.50116324) (356,2.49826036) (357,2.49561363) (358,2.50043348) (359,2.50120361) (360,2.5002113) (361,2.51743127) (362,2.50776044) (363,2.4927285) (364,2.51254486) (365,2.49770534) (366,2.47804897) (367,2.50834656) (368,2.47909085) (369,2.50630954) (370,2.48714527) (371,2.50991343) (372,2.48827002) (373,2.48878825) (374,2.48344094) (375,2.4977489) (376,2.47714351) (377,2.50160244) (378,2.47957552) (379,2.50687936) (380,2.48892062) (381,2.50719929) (382,2.47986907) (383,2.51546331) (384,2.48420505) (385,2.51298786) (386,2.49029148) (387,2.49600201) (388,2.48930972) (389,2.51035438) (390,2.49349392) (391,2.47083885) (392,2.49865069) (393,2.47341726) (394,2.50003581) (395,2.48792347) (396,2.50721153) (397,2.48735503) (398,2.50224488) (399,2.49484189) (400,2.50120817) (404,2.50084703) (408,2.51267543) (412,2.51718159) (416,2.4795095) (420,2.49625575) (424,2.50668393) (428,2.48131614) (432,2.49481281) (436,2.49369488) (440,2.50893482) (444,2.53015647) (448,2.54152162) (452,2.5454856) (456,2.52163526) (460,2.53477547) (464,2.54632024) (468,2.56037842) (472,2.54639333) (476,2.55509993) (480,2.56265109) (484,2.54761235) (488,2.53587946) (492,2.54621281) (496,2.55871649) (500,2.5659789) (504,2.57555405) (508,2.57473533) (512,2.58320184) (516,2.5525364) (520,2.56172395) (524,2.57228595) (528,2.55194282) (532,2.56319382) (536,2.56673052) (540,2.56938863) (544,2.5818622) (548,2.55161876) (552,2.55573179) (556,2.56161947) (560,2.56047186) (564,2.56269094) (568,2.54857875) (572,2.55186025) (576,2.5620554) (580,2.54503768) (584,2.55900803) (588,2.53862922) (592,2.54654418) (596,2.55620229) (600,2.55780783) (604,2.56771829) (608,2.56561453) (612,2.54072714) (616,2.54098149) (620,2.54321626) (624,2.54644158) (628,2.54476261) (632,2.55209563) (636,2.5531251) (640,2.55736798) (644,2.54837937) (648,2.5519119) (652,2.52474767) (656,2.52789645) (660,2.53637468) (664,2.53718976) (668,2.54204807) (672,2.54106796) (676,2.55574902) (680,2.5550764) (684,2.56476328) (688,2.56421512) (692,2.56840173) (696,2.56764644) (700,2.56933757) (704,2.57426276) (708,2.57594583) (712,2.55528995) (716,2.5333367) (720,2.54195829) (724,2.54407957) (728,2.54995548) (732,2.55377589) (736,2.56080354) (740,2.54864679) (744,2.55380045) (748,2.54629529) (752,2.55546568) (756,2.55388888) (760,2.5570744) (764,2.55144189) (768,2.55708692) (772,2.55877447) (776,2.56393166) (780,2.56939538) (784,2.57946601) (788,2.54492759) (792,2.55328255) (796,2.56516919) (800,2.57034666) (804,2.57155649) (808,2.57835337) (812,2.57587831) (816,2.58145958) (820,2.58484354) (824,2.56144386) (828,2.56559345) (832,2.57180932) (836,2.56577931) (840,2.57429331) (844,2.57937267) (848,2.58777386) (852,2.58974933) (856,2.59783118) (860,2.57883577) (864,2.57216798) (868,2.5743022) (872,2.58131724) (876,2.58812469) (880,2.59272915) (884,2.58283519) (888,2.58939338) (892,2.59254949) (896,2.58989294) (900,2.58192392) (904,2.58189272) (908,2.59141071) (912,2.58657688) (916,2.59251649) (920,2.59355605) (924,2.59363056) (928,2.59737525) (932,2.60136271) (936,2.57979791) (940,2.58590458) (944,2.56057111) (948,2.56737957) (952,2.56516413) (956,2.57049199) (960,2.56399241) (964,2.5716145) (968,2.57376945) (972,2.5736957) (976,2.58112679) (980,2.58288319) (984,2.56714123) (988,2.56390193) (992,2.56779059) (996,2.56979894) (1000,2.57481181) (1004,2.57917633) (1008,2.58002137) (1012,2.58649285) (1016,2.56469134) (1020,2.57066034) (1024,2.5717878) (1028,2.56189368) (1032,2.56503448) (1036,2.57056893) (1040,2.57543551) (1044,2.57111513) (1048,2.57517056) (1052,2.57866429) (1056,2.58539721) (1060,2.58892458) (1064,2.58848476) (1068,2.55843105) (1072,2.5639747) (1076,2.57021428) (1080,2.56843338) (1084,2.57399) (1088,2.57665943) (1092,2.58547195) (1096,2.5698644) (1100,2.57536154) (1104,2.58175066) (1108,2.58547569) (1112,2.56984029) (1116,2.57490012) (1120,2.58041442) (1124,2.58564221) (1128,2.58582954) (1132,2.58669417) (1136,2.58950147) (1140,2.59653955) (1144,2.59922993) (1148,2.60459584) (1152,2.60876458) (1156,2.60337699) (1160,2.60282494) (1164,2.6085912) (1168,2.61303871) (1172,2.61524586) (1176,2.61681802) (1180,2.62385637) (1184,2.59945657) (1188,2.60638008) (1192,2.61044106) (1196,2.61449243) (1200,2.6169186) (1204,2.59505274) (1208,2.6020372) (1212,2.59588846) (1216,2.59841854) (1220,2.60139878) (1224,2.60506794) (1228,2.61096851) (1232,2.6097738) (1236,2.61154988) (1240,2.61759542) (1244,2.61100309) (1248,2.61335674) (1252,2.59928281) (1256,2.60063524) (1260,2.60325442) (1264,2.60390461) (1268,2.60908032) (1272,2.59243952) (1276,2.59627579) (1280,2.59938836) (1284,2.60277192) (1288,2.60672604) (1292,2.60968176) (1296,2.61272088) (1300,2.60755583) (1304,2.60237629) (1308,2.60396074) (1312,2.60507274) (1316,2.61107688) (1320,2.61000867) (1324,2.60851212) (1328,2.61216976) (1332,2.61437838) (1336,2.61111743) (1340,2.6135655) (1344,2.61694382) (1348,2.59711194) (1352,2.59985026) (1356,2.60285958) (1360,2.60440723) (1364,2.58534987) (1368,2.58824213) (1372,2.59465793) (1376,2.59544271) (1380,2.59819847) (1384,2.60010681) (1388,2.60139268) (1392,2.60431792) (1396,2.58293246) (1400,2.58772739) (1404,2.58888631) (1408,2.59190382) (1412,2.59632797) (1416,2.5991756) (1420,2.60065778) (1424,2.59954605) (1428,2.59904476) (1432,2.59884823) (1436,2.60277752) (1440,2.6058063) (1444,2.61132223) (1448,2.6060874) (1452,2.60875053) (1456,2.61192331) (1460,2.59540497) (1464,2.59807439) (1468,2.60141519) (1472,2.60241254) (1476,2.60448667) (1480,2.60558077) (1484,2.60699857) (1488,2.60804993) (1492,2.60274586) (1496,2.59002874) (1500,2.59044528) (1504,2.59211859) (1508,2.59330347) (1512,2.59334839) (1516,2.59477394) (1520,2.59463189) (1524,2.59888853) (1528,2.59459646) (1532,2.59655177) (1536,2.59568189) (1540,2.59844789) (1544,2.60270794) (1548,2.59508113) (1552,2.59884086) (1556,2.6034405) (1560,2.58690433) (1564,2.59030936) (1568,2.59545723) (1572,2.59798316) (1576,2.5991979) (1580,2.60205055) (1584,2.60308022) (1588,2.60276205) (1592,2.60072185) (1596,2.60432469) (1600,2.60649435) (1604,2.59861557) (1608,2.60361513) (1612,2.60771657) (1616,2.60831976) (1620,2.61033841) (1624,2.61291631) (1628,2.61308978) (1632,2.6139059) (1636,2.618264) (1640,2.61946345) (1644,2.62413498) (1648,2.62651329) (1652,2.63220065) (1656,2.6094896) (1660,2.61462812) (1664,2.597504) (1668,2.601888) (1672,2.60302323) (1676,2.6055408) (1680,2.60979717) (1684,2.61309154) (1688,2.61684473) (1692,2.62097011) (1696,2.61188748) (1700,2.61626861) (1704,2.616258) (1708,2.61989658) (1712,2.61193693) (1716,2.61654914) (1720,2.61537405) (1724,2.61992473) (1728,2.62171103) (1732,2.61852704) (1736,2.62063887) (1740,2.62643953) (1744,2.62617937) (1748,2.62920502) (1752,2.6312358) (1756,2.63333029) (1760,2.63581372) (1764,2.63523794) (1768,2.60878753) (1772,2.61402228) (1776,2.61716314) (1780,2.62088812) (1784,2.62471834) (1788,2.62751865) (1792,2.62723588) (1796,2.63181588) (1800,2.63086565) (1804,2.62399879) (1808,2.6285004) (1812,2.62999986) (1816,2.63076188) (1820,2.63478841) (1824,2.61010932) (1828,2.61285205) (1832,2.61535019) (1836,2.61977854) (1840,2.6150191) (1844,2.61496162) (1848,2.62005492) (1852,2.62098454) (1856,2.6233312) (1860,2.6244727) (1864,2.62452825) (1868,2.62500559) (1872,2.6282859) (1876,2.61273445) (1880,2.6041546) (1884,2.60604551) (1888,2.60493608) (1892,2.60731534) (1896,2.61013229) (1900,2.61072688) (1904,2.61309127) (1908,2.61601133) (1912,2.60803682) (1916,2.61142095) (1920,2.61230031) (1924,2.61575703) (1928,2.61570585) (1932,2.62025267) (1936,2.62152514) (1940,2.61086291) (1944,2.61486114) (1948,2.6183106) (1952,2.61418189) (1956,2.61919128) (1960,2.62104268) (1964,2.62022313) (1968,2.62497049) (1972,2.6269915) (1976,2.62960162) (1980,2.6318762) (1984,2.63552833) (1988,2.62122048) (1992,2.62459582) (1996,2.60034584) (2000,2.6027141) (2050,2.62331921) (2183,2.62838757) (2325,2.64215641) (2476,2.62831123) (2637,2.63531695) (2808,2.63775544) (2990,2.6476838) (3185,2.63782646) (3391,2.65420433) (3612,2.65420532) (3846,2.65285964) (3956,2.64777153) (4096,2.65457503)};
\addlegendentry{raw sequential state, tie $+$}
\addplot[orange!80!black,only marks,mark=triangle,mark size=1.6pt] coordinates {(4,2.10815763) (9,2.59213451) (16,2.41715084) (25,2.38139796) (36,2.26576763) (49,2.33731177) (64,2.29560987) (81,2.40058156) (100,2.41113647) (121,2.42249851) (144,2.38287865) (169,2.42268803) (196,2.42375435) (225,2.47195478) (256,2.46195583) (289,2.48594925) (324,2.44378028) (361,2.47385544) (400,2.4552581) (484,2.51665284) (576,2.5258514) (676,2.50471148) (784,2.54201327) (900,2.55552133) (1024,2.54272194) (1156,2.56543995) (1296,2.58533352) (1444,2.57907457) (1600,2.57864539) (1764,2.607475) (1936,2.5950927) (4096,2.63286844)};
\addlegendentry{tie $-$}
\addplot[black,dotted,no marks] coordinates {(4,2.82842712) (4096,2.82842712)};
\addlegendentry{$2\sqrt2$ (exact singlet)}
\addplot[gray,dashed,no marks] coordinates {(4,2) (4096,2)};
\addlegendentry{local bound $2$}
\end{axis}
\end{tikzpicture}
\caption{Bob-optimised CHSH value of the state prepared by the raw sequential protocol from two vacua, Frenkel ladder, $r=1$, for the ranks of Fig.~\ref{fig:scan} up to $4096$ (triangles: $-1$ convention at perfect squares). The corrected protocol gives $2\sqrt2$ (dotted) at every rank with $p>0$, by construction; the raw state approaches it as slowly as the binned defect, and is still $0.26$ below it at $S=1024$ and $0.15$ below it at $S=8000$ (Table~\ref{tab:scan}). The dashed line is the local-realistic bound $2$; by \eqref{eq:selftest}, the value at $S=1024$ certifies a singlet fidelity of at least $0.822$.}
\label{fig:bell}
\end{figure}

\paragraph{The raw protocol in Bell units.}
The two protocols differ in what they prepare, not only in how often they succeed. The corrected protocol prepares, whenever $p>0$, an exact singlet of an exact Pauli pair, whose CHSH value is $2\sqrt2$ identically. The raw sequential protocol (Appendix~\ref{app:rank4}: measure $Z\otimes Z$, then $X\otimes X$, keep both ``$-$'' outcomes) succeeds more often -- $p_{\rm raw}$ lies between $2.6\%$ (at $S=13$) and $13.5\%$ (at $S=58$) over the ranks of Fig.~\ref{fig:scan}(b), and above $6.9\%$ for the sampled ranks $S\ge256$ -- but, because $X$ and $Z$ do not anticommute, its conditional state is not a singlet, and the question is how far from one it is. Figure~\ref{fig:bell} answers this in the quantity a Bell experiment would measure: the CHSH value of the raw state with Alice's settings $X,Z$ and Bob's settings optimised (the trace-norm expression of Appendix~\ref{app:rank4}). It is $2.36$ at $S=8$, $2.29$ at $S=32$, $2.46$ at $S=128$, $2.58$ at $S=512$, $2.57$ at $S=1024$ and $2.68$ at $S=8000$ (Table~\ref{tab:scan}); over the scan it lies between $2.09$ (at $S=17$) and $2.81$, with a median that rises from $2.39$ for $S<16$ to $2.61$ for $1024\le S\le4096$. It is always above the local bound $2$ but still $0.15$ below $2\sqrt2$ at $S=8000$, and it approaches $2\sqrt2$ with the same slow, jittery convergence as the binned defect in Fig.~\ref{fig:scan}(a), of which it is the operational expression. The Bell value also yields a model-independent statement about the state. By the analytic self-testing bound of Ref.~\cite{Kaniewski2016}, any state reaching the CHSH value $\mathcal B$ has singlet extractability -- the fidelity to a two-qubit singlet attainable by local isometries that may discard auxiliary degrees of freedom, not the overlap with a particular singlet vector in the original space -- at least
\begin{equation}
F\;\ge\;\max\!\left\{\frac12,\,\frac12+\frac12\,\frac{\mathcal B-\mathcal B^*}{2\sqrt2-\mathcal B^*}\right\},\qquad \mathcal B^*=\frac{16+14\sqrt2}{17}=2.1058,
\label{eq:selftest}
\end{equation}
equal to the trivial bound $1/2$ below $\mathcal B^*$ and to $1$ at $2\sqrt2$. Applied to the raw state this certifies a singlet fidelity of at least $0.676$ at $S=8$, $0.625$ at $32$, $0.748$ at $128$, $0.830$ at $512$, $0.822$ at $1024$, $0.874$ at $3956$ and $0.896$ at $8000$ (bounds rounded down); at $S=17$, the worst rank of the scan, the bound gives nothing beyond the trivial value $1/2$ (it is a bound, not evidence against entanglement), and for every sampled rank $S\ge64$ with the $+1$ edge convention it is at least $0.64$ (the minimum, $0.642$, is at $S=71$; with the $-1$ convention the $S=64$ value is $0.631$). The comparison between the two protocols is thus: a higher success probability with a state whose certified singlet fidelity lags $1$ by an amount set by the quadrature lattice, against a lower success probability with a state whose singlet fidelity is exactly $1$. Both converge to the oscillator values (Sec.~\ref{sec:limit}).

\paragraph{Scale and cutoff.}
For the $S=8$ vacuum the retained even-sector eigenvalues of $K$ are $1/8$ and $-3/32$, so every $\kappa\in[1/8,1)$ gives the same probability (odd components of coherent initial states can enter at larger cutoffs), and for $\kappa<3/32$ nothing is retained. At $S=32$, raising $\kappa$ from $0.5$ to $0.9$ raises the vacuum probability from $0.0070$ to $0.0180$. At $S=8$ the balanced scale that minimises the leading displacement defect gives $p=0$, while $r=0.8$ and $1.6$ give $0.0137$ and $0.0264$: optimising the drive (A) and optimising the homodyne task (B) are different problems.

\paragraph{Lattice points on bin edges are not a convention problem but a resolution problem.}
The two uniform sign assignments at exact edges do not coincide with the one-sided limits in the scale $r$, because successive edges have alternating signs; a slightly detuned homodyne phase or scale moves the edge atoms into different bins. Table~\ref{tab:robust} (Appendix~\ref{app:num}) shows this with $b=\pi/(2r)$ enforced: the $S=8$ vacuum probability is unchanged by $\delta r=\pm10^{-8}$, whereas at $S=16$ it changes from $2\times10^{-5}$ to $5\times10^{-3}$ on either side. Proposition~\ref{prop:margin} and \eqref{eq:smoothbounds} quantify the stable regime and the limitations of smoothing at a crossing.

\section{Wannier-like spectra: compositeness that does not go away}
\label{sec:purity}
For a Frenkel-like coboson the composite nature disappears as the aggregate grows. For a Wannier exciton, a dimer, or any composite whose Schmidt distribution is localised over an infinite mode set, it does not: adding modes leaves the purity, and with it the commutator and the Weyl defect, at a nonzero limit, and there is no oscillator limit for any detector to converge to. The question is then not how fast the composite becomes bosonic but whether a qubit can be extracted from it at all, and at what efficiency. We answer it for two model spectra. The first is the geometric class of Ref.~\cite{Law2005},
\begin{equation}
\lambda_k^{(S)}=\frac{(1-z)z^{k-1}}{1-z^S},\quad
P_S=\frac{1-z}{1+z}\frac{1+z^S}{1-z^S},\quad
P_\infty=\frac{1-z}{1+z},\qquad 0<z<1.
\label{eq:geom}
\end{equation}
In Ref.~\cite{Law2005} this spectrum arises from a two-particle Gaussian wavefunction; we use it as a model Schmidt distribution without identifying it with a specific exciton. At fixed $z$ the purity has a nonzero limit as modes are added, so there is no oscillator limit; the limit $z\to1$ is a different one. The omitted weight before renormalisation is $z^S$; since the binned observables and the spectral filter are discontinuous, a small omitted weight does not by itself bound their change.

The full paired space has dimension $2^S$, with an invariant even-parity vacuum sector. Product Hadamard transforms apply $X$ and $Z$ in $O(S2^S)$ operations. For $S\le12$ the even sector is diagonalised densely (and, as a check, at $S=13$ and $14$ for $z=0.7$); larger calculations use a Lanczos approximation to functions of $K$ (Appendix~\ref{app:num}), and we keep the two regimes apart.

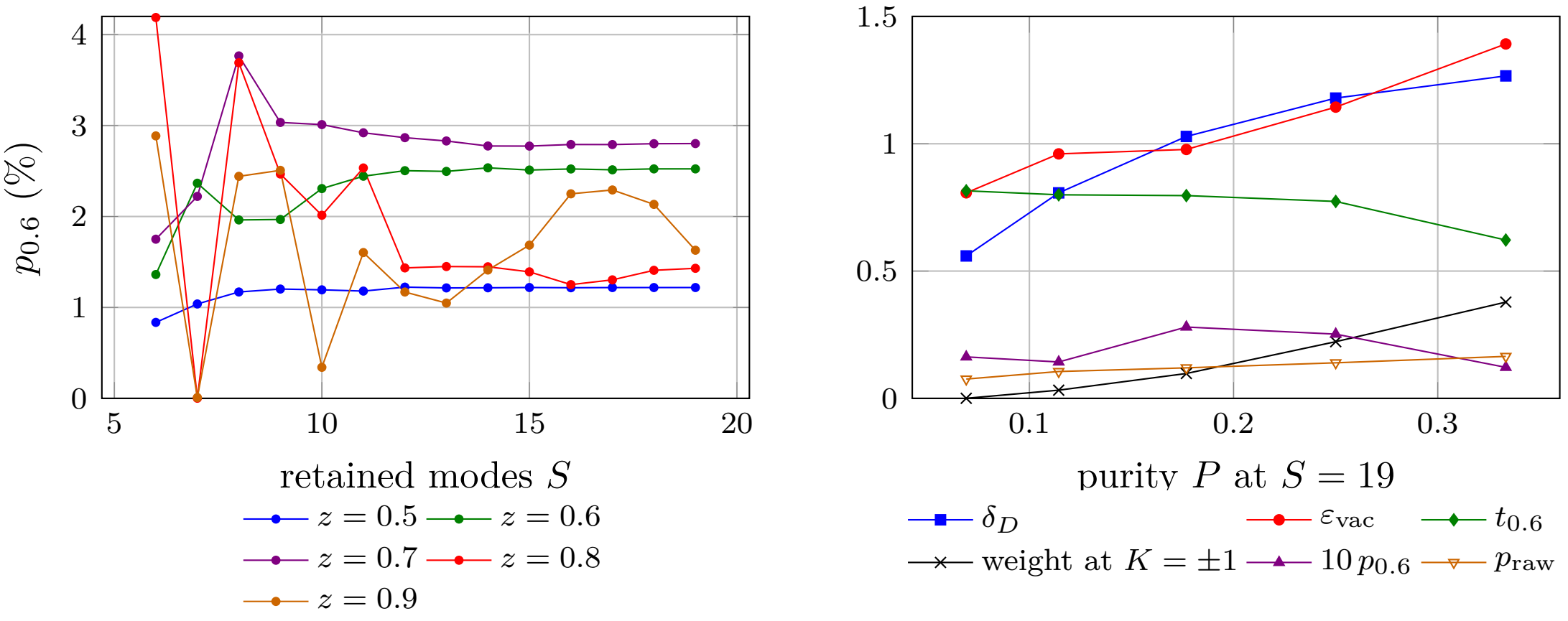
\begin{figure}[t]
\centering
\begin{tikzpicture}
\begin{groupplot}[group style={group size=2 by 1,horizontal sep=1.5cm},width=0.46\textwidth,height=5.1cm,grid=major,tick label style={font=\scriptsize},label style={font=\small},legend style={font=\scriptsize,at={(0.5,-0.24)},anchor=north,draw=none},legend cell align=left,legend columns=2]
\nextgroupplot[xlabel={retained modes $S$},ylabel={$p_{0.6}$ (\%)},ymin=0,ymax=4.2]
\addplot[blue,mark=*,mark size=1pt] coordinates {(6,0.83555065) (7,1.03815512) (8,1.1699175) (9,1.20219444) (10,1.19353623) (11,1.17980389) (12,1.22321688) (13,1.21467552) (14,1.21594313) (15,1.21909292) (16,1.21658736) (17,1.21844318) (18,1.21860748) (19,1.21882715)};
\addlegendentry{$z=0.5$}
\addplot[green!50!black,mark=*,mark size=1pt] coordinates {(6,1.3620164) (7,2.36594227) (8,1.96227589) (9,1.96709422) (10,2.30698893) (11,2.44302147) (12,2.5038285) (13,2.49588951) (14,2.53506152) (15,2.51098067) (16,2.52264199) (17,2.51339369) (18,2.52354957) (19,2.52351025)};
\addlegendentry{$z=0.6$}
\addplot[violet,mark=*,mark size=1pt] coordinates {(6,1.7504349) (7,2.22101475) (8,3.76585824) (9,3.0342906) (10,3.01064994) (11,2.92058476) (12,2.86676344) (13,2.83051523) (14,2.77516569) (15,2.77422926) (16,2.79168878) (17,2.7911099) (18,2.80051073) (19,2.80193258)};
\addlegendentry{$z=0.7$}
\addplot[red,mark=*,mark size=1pt] coordinates {(6,4.18843545) (7,0.00159968552) (8,3.69071001) (9,2.46704988) (10,2.01360844) (11,2.53369553) (12,1.43467447) (13,1.44976572) (14,1.44710156) (15,1.39136603) (16,1.25063108) (17,1.30320352) (18,1.40823741) (19,1.43022991)};
\addlegendentry{$z=0.8$}
\addplot[orange!80!black,mark=*,mark size=1pt] coordinates {(6,2.88635212) (7,0.0092145404) (8,2.44163577) (9,2.50754595) (10,0.341369007) (11,1.60356381) (12,1.16943493) (13,1.0480039) (14,1.41140599) (15,1.68475848) (16,2.24927682) (17,2.29178266) (18,2.13333205) (19,1.6290066)};
\addlegendentry{$z=0.9$}
\nextgroupplot[xlabel={purity $P$ at $S=19$},ymin=0,ymax=1.5,legend columns=3]
\addplot[blue,mark=square*,mark size=1.3pt] coordinates {(0.0690719165,0.559397606) (0.114360499,0.80702283) (0.176873361,1.0285813) (0.25003047,1.17929598) (0.333334605,1.26642893)};
\addlegendentry{$\delta_D$}
\addplot[red,mark=*,mark size=1.3pt] coordinates {(0.0690719165,0.807188178) (0.114360499,0.960162418) (0.176873361,0.977579506) (0.25003047,1.14382295) (0.333334605,1.39196128)};
\addlegendentry{$\varepsilon_{\rm vac}$}
\addplot[green!50!black,mark=diamond*,mark size=1.5pt] coordinates {(0.0690719165,0.815167908) (0.114360499,0.799518057) (0.176873361,0.796354741) (0.25003047,0.773424715) (0.333334605,0.621893199)};
\addlegendentry{$t_{0.6}$}
\addplot[black,mark=x,mark size=2pt] coordinates {(0.0690719165,0.000337468872) (0.114360499,0.0322185059) (0.176873361,0.0976767041) (0.25003047,0.222634252) (0.333334605,0.378106801)};
\addlegendentry{weight at $K=\pm1$}
\addplot[violet,mark=triangle*,mark size=1.5pt] coordinates {(0.0690719165,0.16290066) (0.114360499,0.143022991) (0.176873361,0.280193258) (0.25003047,0.252351025) (0.333334605,0.121882715)};
\addlegendentry{$10\,p_{0.6}$}
\addplot[orange!80!black,mark=triangle,mark size=1.5pt,mark options={rotate=180}] coordinates {(0.0690719165,0.0758819694) (0.114360499,0.105515985) (0.176873361,0.119925836) (0.25003047,0.139705099) (0.333334605,0.164890033)};
\addlegendentry{$p_{\rm raw}$}
\end{groupplot}
\end{tikzpicture}
\caption{Wannier-like (geometric) spectra at $r=1$: the composite stays composite as modes are added, and the singlet probability remains at the percent level over the computed truncations. Left: corrected-protocol vacuum probability at $\kappa=0.6$ versus the number of retained modes, for Schmidt numbers $\mathcal K_\infty=3$ ($z=0.5$) to $19$ ($z=0.9$); $S\le12$ dense, $S\ge13$ Lanczos. For $z\le0.6$ the curves change little over $13\le S\le19$; the $S=19$ truncations have Weyl defects $1.27$ and $1.18$ and singlet probabilities $1.2\%$ and $2.5\%$, consistent with nonzero limits (Proposition~\ref{prop:infinite} and the discussion below). For $z=0.9$ the omitted weight is still $13\%$ at $S=19$ and the curve has not settled. Right: the values at $S=19$ versus the purity $P$ (from $z=0.9$ at the left to $z=0.5$ at the right). As the composite becomes less bosonic both defects rise to order one, the vacuum weight on the one-dimensional Jordan blocks -- which carry no qubit -- rises from $0$ to $0.38$ and becomes the dominant loss, the accepted weight $t_{0.6}$ falls, and the singlet probability $p_{0.6}$ stays at the percent level throughout; $p_{\rm raw}$ is the raw sequential value, not a singlet probability. The two leftmost points ($z=0.8$, $0.9$; omitted weights $0.014$ and $0.135$) are finite-mode values that are still changing with $S$.}
\label{fig:purity}
\end{figure}

\paragraph{Checked finite-mode examples.}
Table~\ref{tab:dense} gives fresh dense calculations for geometric spectra and for the algebraically decaying model
\begin{equation}
\lambda_n^{(S)}\propto[1+(n/w)^2]^{-4},\qquad n=1,\ldots,S.
\label{eq:algebraic}
\end{equation}
The exponent is that of the hydrogenic $1s$ momentum density, but the sequence carries no three-dimensional momentum grid, angular degeneracy or normalisation volume, so it is a model of a Wannier-like tail rather than a derived exciton spectrum. Both families give finite-mode examples with $p>0$ although the displacement and binned defects are of order one. For a sufficiently concentrated spectrum one instead obtains $X=I$: at $w=2$, $S=19$, $\sum_n\sqrt{\lambda_n}<\pi/2$, so all $Q_r$ eigenvalues at $r=1$ lie in the central bin, $K=Z$, and no sector with $|K|<1$ remains.

\begin{table}[t]
\centering\small
\begin{tabular}{lrrrrrr}
\toprule
model & $S$ & $P$ & $\delta_D$ & $\varepsilon_0$ & $p_{0.4}$ & $p_{0.6}$\\
\midrule
geometric $z=0.7$ & 8 & 0.1981 & 1.089 & 0.997 & 0.020650 & 0.037659\\
geometric $z=0.5$ & 10 & 0.3340 & 1.267 & 1.393 & 0.002998 & 0.011935\\
geometric $z=0.7$ & 10 & 0.1867 & 1.058 & 0.978 & 0.020780 & 0.030106\\
algebraic $w=3$ & 10 & 0.5153 & 1.303 & 1.631 & 0.000470 & 0.000676\\
algebraic $w=4$ & 10 & 0.3813 & 1.302 & 1.513 & 0.001535 & 0.005982\\
algebraic $w=6$ & 10 & 0.2480 & 1.189 & 1.156 & 0.005082 & 0.021036\\
\bottomrule
\end{tabular}
\caption{Singlet extraction despite order-one Weyl and binned defects: finite-mode vacuum examples at $r=1$ from dense diagonalisation of the even sector, at cutoffs $\kappa=0.4$ and $0.6$. All have Weyl defects and binned defects of order one; at $\kappa=0.6$ four of the six yield a singlet with probability above $1\%$, the two most concentrated algebraic spectra ($w=3$, $\mathcal K=1.9$; $w=4$, $\mathcal K=2.6$) below it.}
\label{tab:dense}
\end{table}

\paragraph{Filtering and initial-state structure.}
For these spectra $K$ has, besides a quasi-continuous part, eigenvalues at $\pm1$ and -- at $z\le0.7$ -- at $-1/2$ (to $10^{-14}$ in the dense calculations), with large vacuum weight: at $z=0.5$, $S=10$, the dense calculation gives weight $0.378$ at $\pm1$ and $0.187$ at $-1/2$; at $z=0.7$, $S=8$, the weight at $-1/2$ is $0.225$. The weight at $\pm1$ sits on one-dimensional Jordan blocks, carries no qubit and must be rejected; it is the dominant loss at small Schmidt number. We have not derived the eigenvalue $-1/2$ analytically, and whether these atoms survive the infinite-mode limit is not known; we use the cutoffs $0.4$ and $0.6$ to stay away from it. The accepted weight is fixed by the spectral measure of $K$, whereas the conditional probability depends in addition on the reduced qubit state, Eq.~\eqref{eq:mixed}.

\begin{figure}[t]
\centering
\definecolor{cutblue}{RGB}{0,114,178}
\definecolor{cutorange}{RGB}{230,159,0}
\definecolor{cutgreen}{RGB}{0,158,115}
\definecolor{cutred}{RGB}{213,94,0}
\definecolor{cutpurple}{RGB}{160,88,153}
\begin{tikzpicture}
\begin{groupplot}[group style={group size=1 by 2,vertical sep=1.25cm},width=.87\textwidth,xmin=0,xmax=1,xtick={0,0.2,0.4,0.6,0.8,1},grid=major,grid style={gray!18},tick label style={font=\small},label style={font=\small}]
\nextgroupplot[height=5.1cm,ymin=0,ymax=5,ytick={0,1,2,3,4,5},xticklabels=\empty,ylabel={singlet probability (\%)},title={(a) Preparation probability},title style={font=\small,at={(0,1.03)},anchor=south west},legend to name=cutofflegend,legend columns=5,legend style={font=\scriptsize,draw=none,column sep=6pt},legend cell align=left]
\draw[gray!65,dashed] (axis cs:0.4,0)--(axis cs:0.4,5);
\draw[gray!65,dashed] (axis cs:0.6,0)--(axis cs:0.6,5);
\addplot[cutblue,solid,thick,const plot,no marks] coordinates {(0.001,0) (0.0937500001,4.72712147e-15) (0.125,1.24246318) (0.999,1.24246318)};
\addlegendentry{flat $S=8$}
\addplot[cutorange,densely dashed,thick,const plot,no marks] coordinates {(0.001,4.41487125e-14) (0.00236352916,0.012132547) (0.092884455,0.19448763) (0.129662683,0.591225507) (0.255631625,0.689981648) (0.3126369,2.5148515) (0.608160927,2.87296657) (0.66977349,2.88471213) (0.816551006,3.9894242) (0.842847703,4.28861739) (0.884364959,4.43077642) (0.99230402,4.75294975) (0.999,4.75294975)};
\addlegendentry{flat $S=64$}
\addplot[cutgreen,dash dot,thick,const plot,no marks] coordinates {(0.001,0) (0.0655730572,2.92821937e-19) (0.103138525,1.52490552e-05) (0.152402949,0.00327915071) (0.165012,0.00559849309) (0.19968623,0.0174177507) (0.25,0.0174177507) (0.307393247,0.0642832488) (0.375,0.184909581) (0.410097051,0.273062805) (0.445497991,0.876382745) (0.5,1.1699175) (0.999,1.1699175)};
\addlegendentry{geometric $z=0.5$}
\addplot[cutred,dashed,thick,const plot,no marks] coordinates {(0.001,0) (0.117256831,0) (0.125,0.529910327) (0.25,1.22657491) (0.308718501,1.35895075) (0.340720517,2.06503041) (0.5,3.76585824) (0.811567813,3.82684953) (0.999,3.82684953)};
\addlegendentry{geometric $z=0.7$}
\addplot[cutpurple,densely dotted,thick,const plot,no marks] coordinates {(0.001,0) (0.0333793874,0) (0.0926684973,0.001708473) (0.12100237,0.0128743844) (0.157709907,0.0384077477) (0.25,0.0514158879) (0.307042463,0.149795909) (0.31571089,0.336468662) (0.375,0.336468662) (0.5,2.48269665) (0.999,2.48269665)};
\addlegendentry{algebraic $w=6$}
\nextgroupplot[height=3.8cm,ymin=0,ymax=1.5,ytick={0,0.5,1,1.41421356237},yticklabels={$0$,$0.5$,$1$,$\sqrt2$},xlabel={filter cutoff $\kappa$},ylabel={bound $C(\kappa)$},title={(b) Allowed measurement correction},title style={font=\small,at={(0,1.03)},anchor=south west}]
\draw[gray!65,dashed] (axis cs:0.4,0)--(axis cs:0.4,1.5);
\draw[gray!65,dashed] (axis cs:0.6,0)--(axis cs:0.6,1.5);
\addplot[black,thick,no marks] coordinates {(0,0) (0.00785390089,0.00785396145) (0.0157073173,0.0157078018) (0.0235597648,0.0235613999) (0.0314107591,0.0314146346) (0.0392598158,0.0392673849) (0.0471064507,0.0471195297) (0.0549501799,0.0549709478) (0.0627905195,0.0628215182) (0.0706269859,0.0706711198) (0.0784590957,0.0785196315) (0.0862863658,0.0863669324) (0.0941083133,0.0942129014) (0.101924456,0.102057418) (0.109734311,0.10990036) (0.117537397,0.117741607) (0.125333234,0.125581039) (0.133121339,0.133418534) (0.140901232,0.141253972) (0.148672434,0.149087231) (0.156434465,0.156918191) (0.164186847,0.164746732) (0.1719291,0.172572732) (0.179660749,0.18039607) (0.187381315,0.188216627) (0.195090322,0.196034281) (0.202787295,0.203848912) (0.21047176,0.211660399) (0.218143241,0.219468622) (0.225801267,0.227273461) (0.233445364,0.235074795) (0.241075061,0.242872504) (0.248689887,0.250666467) (0.256289373,0.258456565) (0.26387305,0.266242677) (0.27144045,0.274024683) (0.278991106,0.281802464) (0.286524553,0.289575899) (0.294040325,0.297344868) (0.30153796,0.305109252) (0.309016994,0.31286893) (0.316476967,0.320623784) (0.323917418,0.328373693) (0.331337888,0.336118539) (0.33873792,0.343858201) (0.346117057,0.35159256) (0.353474844,0.359321497) (0.360810826,0.367044893) (0.368124553,0.374762629) (0.375415571,0.382474586) (0.382683432,0.390180644) (0.389927688,0.397880685) (0.397147891,0.405574591) (0.404343596,0.413262242) (0.411514359,0.42094352) (0.418659738,0.428618306) (0.425779292,0.436286483) (0.432872582,0.443947931) (0.43993917,0.451602534) (0.446978621,0.459250172) (0.4539905,0.466890728) (0.460974375,0.474524084) (0.467929814,0.482150122) (0.47485639,0.489768724) (0.481753674,0.497379774) (0.488621241,0.504983154) (0.495458668,0.512578746) (0.502265533,0.520166434) (0.509041416,0.5277461) (0.515785898,0.535317627) (0.522498565,0.5428809) (0.529179001,0.5504358) (0.535826795,0.557982212) (0.542441537,0.565520019) (0.549022818,0.573049105) (0.555570233,0.580569355) (0.562083378,0.58808065) (0.568561851,0.595582877) (0.575005252,0.60307592) (0.581413184,0.610559662) (0.587785252,0.618033989) (0.594121063,0.625498785) (0.600420225,0.632953934) (0.606682351,0.640399323) (0.612907054,0.647834836) (0.619093949,0.655260359) (0.625242656,0.662675777) (0.631352795,0.670080975) (0.63742399,0.67747584) (0.643455865,0.684860258) (0.649448048,0.692234114) (0.655400171,0.699597295) (0.661311865,0.706949688) (0.667182767,0.714291178) (0.673012514,0.721621653) (0.678800746,0.728941) (0.684547106,0.736249105) (0.69025124,0.743545857) (0.695912797,0.750831142) (0.701531426,0.758104849) (0.707106781,0.765366865) (0.712638519,0.772617078) (0.718126298,0.779855376) (0.723569779,0.787081647) (0.728968627,0.794295781) (0.734322509,0.801497666) (0.739631095,0.808687191) (0.744894057,0.815864245) (0.75011107,0.823028717) (0.755281812,0.830180497) (0.760405966,0.837319475) (0.765483213,0.84444554) (0.770513243,0.851558583) (0.775495743,0.858658494) (0.780430407,0.865745163) (0.785316931,0.872818481) (0.790155012,0.87987834) (0.794944353,0.886924629) (0.799684658,0.893957241) (0.804375635,0.900976067) (0.809016994,0.907980999) (0.81360845,0.914971929) (0.818149717,0.921948749) (0.822640518,0.928911351) (0.827080574,0.935859629) (0.831469612,0.942793474) (0.835807361,0.94971278) (0.840093554,0.95661744) (0.844327926,0.963507348) (0.848510215,0.970382398) (0.852640164,0.977242483) (0.856717519,0.984087498) (0.860742027,0.990917337) (0.864713441,0.997731895) (0.868631514,1.00453107) (0.872496007,1.01131475) (0.87630668,1.01808283) (0.880063298,1.02483522) (0.88376563,1.0315718) (0.887413447,1.03829247) (0.891006524,1.04499713) (0.89454464,1.05168567) (0.898027576,1.058358) (0.901455117,1.06501401) (0.904827052,1.07165359) (0.908143174,1.07827665) (0.911403277,1.08488307) (0.91460716,1.09147277) (0.917754626,1.09804564) (0.92084548,1.10460157) (0.923879533,1.11114047) (0.926856596,1.11766223) (0.929776486,1.12416676) (0.932639023,1.13065395) (0.935444031,1.1371237) (0.938191336,1.14357592) (0.940880769,1.1500105) (0.943512164,1.15642735) (0.946085359,1.16282637) (0.948600195,1.16920745) (0.951056516,1.1755705) (0.953454172,1.18191543) (0.955793015,1.18824213) (0.958072899,1.1945505) (0.960293686,1.20084045) (0.962455236,1.20711188) (0.964557418,1.2133647) (0.966600102,1.21959881) (0.968583161,1.22581411) (0.970506473,1.2320105) (0.97236992,1.2381879) (0.974173387,1.2443462) (0.975916762,1.25048531) (0.977599938,1.25660514) (0.979222811,1.26270559) (0.98078528,1.26878657) (0.982287251,1.27484798) (0.983728629,1.28088973) (0.985109326,1.28691173) (0.986429257,1.29291388) (0.987688341,1.2988961) (0.988886499,1.30485828) (0.990023658,1.31080034) (0.991099747,1.31672219) (0.992114701,1.32262373) (0.993068457,1.32850488) (0.993960955,1.33436553) (0.994792142,1.34020561) (0.995561965,1.34602503) (0.996270376,1.35182368) (0.996917334,1.35760149) (0.997502796,1.36335836) (0.998026728,1.36909421) (0.998489097,1.37480895) (0.998889875,1.38050248) (0.999229036,1.38617473) (0.99950656,1.39182559) (0.99972243,1.397455) (0.999876632,1.40306285) (0.999969158,1.40864907) (1,1.41421356)};
\addplot[black,only marks,mark=*,mark size=1.8pt] coordinates {(0.4,0.408619287)};
\node[font=\scriptsize,anchor=south east,fill=white,inner sep=1.5pt] at (axis cs:0.4,0.4086192873784399) {$C\simeq0.409$};
\addplot[black,only marks,mark=*,mark size=1.8pt] coordinates {(0.6,0.632455532)};
\node[font=\scriptsize,anchor=south east,fill=white,inner sep=1.5pt] at (axis cs:0.6,0.6324555320336758) {$C\simeq0.632$};
\addplot[black,only marks,mark=o,mark options={fill=white},mark size=2pt] coordinates {(1,1.41421356237)};
\end{groupplot}
\end{tikzpicture}
\par\smallskip\pgfplotslegendfromname{cutofflegend}
\caption{Success probability and allowed measurement correction as the filter is opened, at $r=1$. (a) Total singlet probability from two vacua, including both local filters: flat spectra at $S=8,64$ and nonflat spectra at $S=8$. The dense finite-rank curves are nondecreasing step functions; possible jumps occur at eigenvalues $|k|$ of $|K|$, with $k=\cos\theta$ on a Jordan block. Their heights depend on the initial state. (b) The universal bound $C(\kappa)=\sqrt{2(1-\sqrt{1-\kappa^2})}$ on $\|\widetilde Z-Z\|_E$, not the actual correction for any one model. Dashed guides mark the cutoffs $0.4$ and $0.6$ used in the nonflat tables: the corresponding bounds are approximately $0.409$ and $0.632$, respectively. The two panels share the cutoff axis, so gains in success probability can be read against the allowed change of observable without a second vertical scale. Curves are shown for $0<\kappa<1$; the endpoint in (b) denotes the limiting bound $\sqrt2$.}
\label{fig:cutoff}
\end{figure}
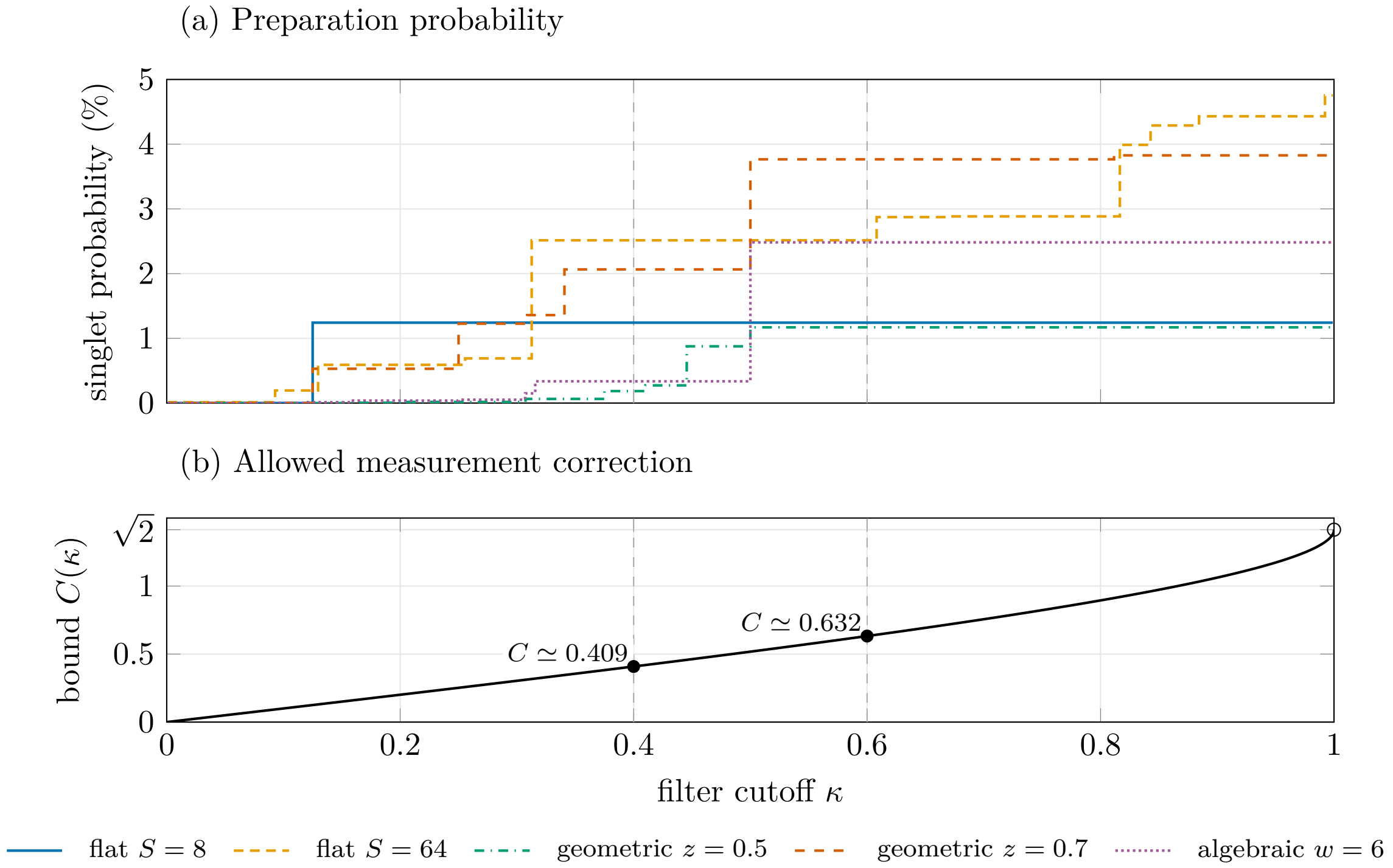

Figure~\ref{fig:cutoff}(a) shows the nondecreasing step functions of Lemma~\ref{lem:cutoff} for dense finite examples. Panel (b) gives the correction bound \eqref{eq:corrbound}, which is $0.30$ at $\kappa=0.3$, $0.52$ at $0.5$ and $0.89$ at $0.8$. The cutoff thus sets success probability against the admissible difference between $\Zt$ and the binned quadrature $Z$; $\kappa=1/2$ elsewhere in this paper is a representative choice, not an optimised one.

\paragraph{Behaviour with the number of modes.}
Figure~\ref{fig:purity} and Table~\ref{tab:purity} give the scan through $S=19$. For $z\le0.6$ the quantities are stable in $S$ once the omitted weight is small: between $S=13$ and $19$, $p_{0.6}$ varies by $4\times10^{-5}$ ($0.4\%$ of its value, between $0.012147$ and $0.012191$) at $z=0.5$ and by $4\times10^{-4}$ ($1.6\%$) at $z=0.6$, $t_{0.6}$ by $6\times10^{-4}$ in both cases, and the $\kappa=0.4$ quantities by approximately $1$--$3.5\%$. This is finite-window stability at the percent level, not a certified value of an infinite-mode limit; the existence of that limit is discussed below. At $z=0.7$ the dependence on $S$ is still visible at the cutoff $0.4$ -- $t_{0.4}$ decreases from $0.653$ at $S=12$ to $0.603$ at $19$ -- but not at $0.6$, where $p_{0.6}$ is $0.0279\pm0.0002$ for $S\ge14$; dense even-sector diagonalisations at $S=13$ and $14$ reproduce the Lanczos values to four digits ($t_{0.4}=0.6245$, $p_{0.4}=0.02072$, $t_{0.6}=0.7838$, $p_{0.6}=0.02831$ at $S=13$; $t_{0.4}=0.6474$, $p_{0.4}=0.02156$, $t_{0.6}=0.7770$, $p_{0.6}=0.02775$ at $S=14$), so the drift at $\kappa=0.4$ is a genuine mode-truncation effect (eigenvalues of $|K|$ close to $0.4$ moving across the cutoff as modes are added) and not a Krylov artefact. At $z=0.8$ and $0.9$ the omitted weights at $S=19$ are $0.014$ and $0.135$, the finite Schmidt numbers $1/P_{19}$ are $8.7$ and $14.5$ rather than the limiting $9$ and $19$, and the probabilities are still changing with $S$ ($p_{0.6}=0.0229$, $0.0213$, $0.0163$ at $S=17,18,19$ for $z=0.9$); these two rows are finite-mode values only.
\begin{table}[t]
\centering\small
\begin{tabular}{lrrrrrrrr}
\toprule
$z$ & $\mathcal K_S$ & $z^S$ & $\delta_D$ & $\varepsilon_0$ & $t_{0.4}$ & $p_{0.4}$ & $t_{0.6}$ & $p_{0.6}$\\
\midrule
0.5 & 3.00 & 0.0000 & 1.266 & 1.392 & 0.302 & 0.0029 & 0.622 & 0.0122\\
0.6 & 4.00 & 0.0001 & 1.179 & 1.144 & 0.533 & 0.0115 & 0.773 & 0.0252\\
0.7 & 5.65 & 0.0011 & 1.029 & 0.978 & 0.603 & 0.0195 & 0.796 & 0.0280\\
0.8 & 8.74 & 0.0144 & 0.807 & 0.960 & 0.525 & 0.0072 & 0.800 & 0.0143\\
0.9 & 14.48 & 0.1351 & 0.559 & 0.807 & 0.771 & 0.0107 & 0.815 & 0.0163\\
\bottomrule
\end{tabular}
\caption{Geometric class at $S=19$ retained modes, $r=1$, vacuum initial state (Lanczos, $240$ steps). $\mathcal K_S=1/P_S$ is the Schmidt number of the truncated, renormalised spectrum. For $z\le0.6$ the preparation probabilities vary by approximately $0.4$--$3.5\%$ over $13\le S\le19$ (see text); the rows $z=0.8$, $0.9$ are finite-mode values that are still changing with $S$.}
\label{tab:purity}
\end{table}

The Lanczos filter has its own error when the vacuum spectral measure of $K$ is quasi-continuous near the cutoff. Table~\ref{tab:lanczos} (Appendix~\ref{app:num}) shows the sensitivity to the Krylov dimension at $z=0.8$, $S=16$: $t_{0.4}$ moves by $2\%$ and $p_{0.4}$ by $3\%$ between $240$ and $960$ steps, while the $\kappa=0.6$ values move by less than $0.5\%$. Ritz values (the eigenvalue approximations produced by Lanczos) are also unreliable for locating atoms: the weight $0.08$ at $-1/2$ found densely at $z=0.7$, $S=13$ and $14$, is not resolved by the $240$-step Lanczos runs at $S\ge17$, so we make no statement about that atom at $S=19$.

\paragraph{The infinite-mode limit: what converges.}
The convergence question has two parts, the bin edges of the quadratures and the filter boundaries $\pm\kappa$ of $K$. The first can be settled in general, for every spectrum with summable Schmidt amplitudes and for every vector, by an elementary argument.
\begin{lemma}
\label{lem:nonatomic}
Let $a_k=\sqrt{\lambda_k}>0$ for infinitely many $k$, with $\sum_ka_k<\infty$, and let $\varepsilon_k$ be independent fair signs. The law of $q=\sum_ka_k\varepsilon_k$ has no atoms.
\end{lemma}
\begin{proof}
Since $a_k\to0$ one can choose a subsequence $b_1>b_2>\dots$ of the $a_k$ with $b_{j+1}<b_j/4$, so that $\sum_{j>m}b_j<b_m/3$. Two sign sequences on this subsequence that first differ at index $m$ give sums differing by at least $2b_m-2\sum_{j>m}b_j>0$; the map from sign sequences to sums is therefore injective on the subsequence, and every value of the subsequence sum has probability zero. The subsequence sum is independent of the sum over the remaining indices, and the convolution of a nonatomic law with any law is nonatomic.
\end{proof}
\begin{proposition}
\label{prop:infinite}
Let $(\lambda_k)_{k\ge1}$ be a normalised infinite sequence with $\lambda_k>0$ and $\sum_k\sqrt{\lambda_k}<\infty$, fix $r>0$ and $b=\pi/(2r)$, and let $\lambda^{(S)}_k=\lambda_k/\sum_{j\le S}\lambda_j$ for $1\le k\le S$ be the renormalised truncations. Let $\mathcal H$ be the incomplete tensor product built on the vacuum\footnote{von Neumann's incomplete tensor product: the closed span of the product vectors that differ from the vacuum product $\bigotimes_k\ket{0_k}$ in finitely many modes, i.e.\ the completion of the space of finite-occupation vectors. It is the natural infinite-mode state space here because the vacuum and every finite-pair state lie in it.}. Then $X_S\to X_\infty$, $Z_S\to Z_\infty$ and $K_S\to K_\infty$ strongly on $\mathcal H$, and for every $\kappa\in(0,1)$ such that $\pm\kappa$ are not eigenvalues of $K_\infty$ the filter, the corrected observables, the two-party singlet projector and the vacuum preparation probability converge to their infinite-mode values. All but countably many $\kappa$ have this property.
\end{proposition}
\begin{proof}
Expand every mode in the eigenbasis of $d_k+d_k^\dagger$ (the local $\sigma_x$). This identifies $\mathcal H$ with $L^2(\{-1,+1\}^{\mathbb N},\mu)$, $\mu$ the fair product measure: the vacuum is the constant function $1$, a finite-occupation basis vector is the Walsh character $\prod_{k\in A}\varepsilon_k$, and $Q_w^{(\infty)}$ is multiplication by $-b\,q(\varepsilon)$ with $q=\sum_ka_k\varepsilon_k$. The spectral measure of any $\psi\in\mathcal H$ is $\nu_\psi(A)=\int\mathbf 1\{-bq(\varepsilon)\in A\}\,|\psi(\varepsilon)|^2\,d\mu$, hence absolutely continuous with respect to the vacuum's, which is the law of $-bq$ and has no atoms by Lemma~\ref{lem:nonatomic}; so no vector has spectral mass on the countable set of bin edges. 

The truncated quadrature is multiplication by $-bq_S(\varepsilon)$ with $q_S=\sum_{k\le S}a^{(S)}_k\varepsilon_k$, and $\sup_\varepsilon|q_S-q|\le\big(\sum_ka_k\big)\big|(\sum_{j\le S}\lambda_j)^{-1/2}-1\big|+\sum_{k>S}a_k\to0$; off the bin edges the sign $F(-bq_S(\varepsilon))$ therefore eventually equals $F(-bq(\varepsilon))$, and dominated convergence gives $Z_S\psi\to Z_\infty\psi$ for every $\psi$. The quadrature $Q_r$ equals $(r/b)\,i^{\hat N}Q_w\,i^{-\hat N}$, with $\hat N$ the pair number: conjugation by the diagonal unitary $i^{\hat N}$ of the occupation basis (a product of local phases, which preserves $\mathcal H$ and is independent of $S$) sends $B^\dagger+B$ to $i(B^\dagger-B)$. Hence the same holds for $X_S$. 

Strong convergence of uniformly bounded operators passes to products, hence to $K_S$. If $\pm\kappa$ are not eigenvalues of $K_\infty$, the spectral projector $\mathbf 1_{[-\kappa,\kappa]}(K_S)$ converges strongly to $\mathbf 1_{[-\kappa,\kappa]}(K_\infty)$, and the bounded Borel function $g_\kappa(k)=\mathbf1_{[-\kappa,\kappa]}(k)/\sqrt{1-k^2}$ (defined as zero outside the retained interval) is continuous except at $\pm\kappa$. Thus $g_\kappa(K_S)$ also converges strongly. Using $\Zt_S=(Z_S-K_SX_S)g_\kappa(K_S)$ gives convergence of the corrected observables, their two-party singlet projectors and $p$. On a separable space $K_\infty$ has at most countably many eigenvalues.
\end{proof}
Proposition~\ref{prop:infinite} applies to both model families ($a_k\propto z^{k/2}$ and $a_n\propto(1+(n/w)^2)^{-2}$). It settles the bin-edge question for every vector and leaves the filter boundary as the remaining convergence condition. What it does not give is a proof that the particular cutoffs $0.4$ and $0.6$ are not eigenvalues of $|K_\infty|$ for the spectra of Table~\ref{tab:purity}, a proof that the limiting probabilities are positive, or a rate. We therefore state the existence and positivity of the infinite-mode values of Table~\ref{tab:purity} as a conjecture supported by the finite-window stability reported above, and note that Proposition~\ref{prop:infinite} guarantees convergence at all but countably many cutoffs. (For the geometric class the vacuum law is the Bernoulli convolution $\nu_{\sqrt z}$\footnote{The Bernoulli convolution $\nu_\rho$ is the law of $\sum_{n\ge0}\varepsilon_n\rho^n$ for fair signs $\varepsilon_n$. A Pisot number is a real algebraic integer greater than $1$ whose other conjugates lie inside the unit disc; a Garsia number is an algebraic integer whose conjugates all lie outside the unit disc and whose norm is $\pm2$, such as $\sqrt2$, which is why $z=1/2$ is covered.}, whose finer properties -- singularity for Pisot parameters \cite{Erdos1939}, absolute continuity for almost every parameter \cite{Solomyak1995,Shmerkin2014} and for Garsia parameters such as $z=1/2$ \cite{Garsia1962} -- are not needed here.) First binned moments are in any case insufficient to identify the algebra: at $z=0.5$, $\langle Z\rangle_0=0.293$, $21\%$ below the oscillator value, while $\langle X\rangle_0=0.749$ is within $3\%$ of it.

\section{Detector (C): coherence without algebra}
\label{sec:initial}

Intensity correlations are widely used to assess coherence in exciton and polariton condensates \cite{Deng2010}. A value $g^{(2)}\simeq1$ is consistent with a coherent field, but does not by itself characterize a coherent state. The quasi-Glauber states of Ref.~\cite{Shiau2017} are the composite states that pass this test best, and they let us ask whether passing it says anything about detector (B). With $\mu=|\alpha|^2$ define
\begin{equation}
\ket{G_\alpha}=\frac1{\sqrt{T_S}}\sum_{N=0}^S\frac{\alpha^N}{\sqrt{N!F_N}}\ket N,\qquad
T_M=\sum_{N=0}^M\frac{\mu^N}{N!F_N},\qquad \mathcal P_N=\frac{\mu^N}{N!F_NT_S}.
\label{eq:qG}
\end{equation}
The finite ladder gives
\begin{equation}
B\ket{G_\alpha}=\alpha\Big(\ket{G_\alpha}-\frac{\alpha^S}{\sqrt{S!F_ST_S}}\ket S\Big),\qquad
\|(B-\alpha)\ket{G_\alpha}\|^2=\mu\,\mathcal P_S ,
\label{eq:qGeig}
\end{equation}
so the state is an exact eigenstate of $B$ only up to the top-rung term, which cannot vanish for finite nonzero $\alpha$ since $B^{S+1}=0$ excludes a nonzero annihilation eigenvalue. Repeated annihilation yields
\begin{equation}
\langle B^{\dagger m}B^m\rangle_G=\mu^m\frac{T_{S-m}}{T_S},\qquad
g^{(2)}_G=\frac{T_ST_{S-2}}{T_{S-1}^2}=\frac{1-\mathcal P_S-\mathcal P_{S-1}}{(1-\mathcal P_S)^2}.
\label{eq:g2G}
\end{equation}
For $S=8$, $\alpha=1$ this gives $g^{(2)}_G=0.9999881\ldots$ while the same state has $\varepsilon_G=1.5715$ at $r=1$: near-unit second-order coherence coexists with an order-one anticommutation defect of the binned quadratures. We do not read this as a surprise -- $g^{(2)}$ is a property of the state, $\varepsilon$ a property of the state and two specific discontinuous observables -- but it makes concrete that (C) does not control (B).

The quasi-Glauber state must be distinguished from the displaced vacuum. For the flat spectrum and $\xi=\rho e^{i\varphi}$ the product of local rotations gives exactly
\begin{equation}
D_B(\xi)\ket0=(\cos a)^S\,e^{\beta B^\dagger}\ket0,\qquad a=\rho/\sqrt S,\quad \beta=\sqrt S\,e^{i\varphi}\tan a,
\label{eq:anderson}
\end{equation}
for $\cos a\ne0$, i.e. the normalised Anderson state \cite{Anderson1958} of Ref.~\cite{Shiau2017}, Eq.~(50). Its number-state coefficients are proportional to $\beta^N\sqrt{F_N/N!}$, whereas \eqref{eq:qG} contains $F_N^{-1/2}$; the distinction persists at finite rank. Appendix~\ref{app:g2} derives the field correlation of the Anderson state and corrects Eq.~(53) of Ref.~\cite{Shiau2017}.

Table~\ref{tab:initial} lists the corrected-protocol probabilities for the three initial states at $S=8$ and $S=32$ with $r=1$, $\kappa=1/2$ and both coherent-family parameters equal to one; both parity sectors are kept and the full initial normalisation is used. The rows of a group do not have equal mean pair number, so the table is not an optimisation at fixed density. At $S=8$ the odd-parity $K$ eigenvalues are $-1,1,-3/4,-3/4$, so the whole odd sector is rejected and the coherent initial states lose their entire odd component; at $S=32$ the three initial states are comparable. Exact radical expressions for the two nonvacuum $S=8$ entries are given in Appendix~\ref{app:eight}.

\begin{table}[t]
\centering
\small
\begin{tabular}{rlcccc}
\toprule
$S$ & initial state & $t$ & $g^{(2)}$ & $\varepsilon_v$ & $p(v)$\\
\midrule
8 & vacuum & 0.969828 & -- & 0.410435 & 0.012424631790\\
8 & quasi-Glauber & 0.321363 & 0.999988135 & 1.571540 & 0.000274534936\\
8 & Anderson & 0.422886 & 0.789053254 & 1.400949 & 0.000000358389\\
32 & vacuum & 0.743321 & -- & 0.993255 & 0.007038319914\\
32 & quasi-Glauber & 0.762299 & 1.000000000 & 0.874936 & 0.008913682982\\
32 & Anderson & 0.757124 & 0.940306901 & 0.880731 & 0.010084343692\\
\bottomrule
\end{tabular}
\caption{Initial state dependence of the corrected singlet protocol, $r=1$, $\kappa=1/2$, $\alpha=\beta=1$. Entries are rounded; the field-correlation ratio is undefined for the vacuum.}
\label{tab:initial}
\end{table}

\section{Oscillator limit and nonzero overlap}
\label{sec:limit}

Everything above concerns finite $S$. We now show that for the Frenkel ladder the whole construction -- binned quadratures, filter, corrected observables and singlet projector -- converges to the oscillator construction of Ref.~\cite{Sorella2026}, and that the oscillator vacuum overlap at which it arrives is strictly positive at every scale. Embed the symmetric ladder in $\ell^2(\mathbb N_0)$ and extend the finite annihilator by zero above $S$. The coefficients in \eqref{eq:ladder} converge to the oscillator coefficients on every finite-number vector. Since these vectors form a core for each limiting quadrature, the finite quadratures converge in the strong resolvent sense\footnote{A core is a dense subspace on which the operator determines its closure. Strong resolvent convergence means $(Q_S+i)^{-1}f\to(Q_\infty+i)^{-1}f$ for every $f$; it implies $g(Q_S)\to g(Q_\infty)$ strongly for every bounded $g$ whose discontinuities carry no spectral weight of $Q_\infty$, which is what the bin edges of $F$ require.}: for a core vector $f$,
\begin{equation}
(Q_S+i)^{-1}(Q_\infty+i)f-f=(Q_S+i)^{-1}(Q_\infty-Q_S)f\longrightarrow0,
\end{equation}
and the resolvent bound with density of the core image extends this to the whole space. Although $F$ is discontinuous, its discontinuity set is countable and has zero spectral measure for the oscillator quadratures, so bounded functional calculus gives $X_S\to X_\infty$ and $Z_S\to Z_\infty$ strongly. The canonical Weyl relation at $2rb=\pi$ makes every odd harmonic of one quadrature anticommute with every odd harmonic of the other; Fej\'er approximants of the square wave are uniformly bounded and contain only odd harmonics, so their strong limits give $\{X_\infty,Z_\infty\}=0$ and $K_S\to0$ strongly. For every fixed $0<\kappa<1$ it follows that $E_S\to I$ and $\Zt_S\to Z_\infty$ strongly, because $\mathbf 1_{[-\kappa,\kappa]}(k)(1-k^2)^{-1/2}$ is bounded and continuous near the limiting spectral point $k=0$. The two-party projectors converge strongly, so the vacuum preparation probabilities converge to the oscillator value.

\begin{proposition}
\label{prop:positive}
Let
\begin{equation}
f(s)=\frac4\pi\sum_{m=0}^\infty\frac{(-1)^m}{2m+1}e^{-(2m+1)^2s^2/2},\qquad
p_\infty(r)=\frac{1-f(r)^2-f(\pi/2r)^2}{4}.
\label{eq:f}
\end{equation}
Then $0<p_\infty(r)<1/4$ for every $0<r<\infty$. In particular $p_\infty(1)=0.0683497058887\ldots$, with $f(1)=0.767545$ and $f(\pi/2)=0.370777$; at the balanced scale $p_\infty(\sqrt{\pi/2})=0.081709$.
\end{proposition}

\begin{proof}
$f(s)$ is the vacuum expectation of $F(Q)$ for a quadrature of variance $s^2$; the series alternates with strictly decreasing terms, so $f(s)>0$ for finite $s>0$. For the real oscillator vacuum, complex conjugation makes the $Y=iXZ$ expectation vanish exactly. Put $x=f(r)$, $z=f(b)$ and $A=xX+zZ$; anticommutation gives $A^2=(x^2+z^2)I$, so the variance of $A$ in the vacuum is $(x^2+z^2)(1-x^2-z^2)\ge0$, i.e. $x^2+z^2\le1$. If equality held, $(A-I)\psi_0=0$ and hence $xX\psi_0=(I-zZ)\psi_0$. In the position representation $\psi_0(q)=\pi^{-1/4}e^{-q^2/2}$ the right-hand side has a nonzero jump at every bin edge, because $z>0$ and the Gaussian never vanishes; the left-hand side is entire, being the Fourier transform of the Gaussian multiplied by the bounded momentum square wave. Equality almost everywhere would force equal one-sided limits at the edges, a contradiction. The upper bound follows from $x,z>0$.
\end{proof}

These values were confirmed by the author of Ref.~\cite{Sorella2026} (see the Acknowledgments); note that at $r=1$ the two expectations are unequal, $f(1)\ne f(\pi/2)$. The corrected vacuum protocol therefore succeeds for all sufficiently large flat-spectrum ranks at every fixed scale and cutoff. The raw sequential state converges to the oscillator singlet as well, and its Bob-optimised CHSH value approaches $2\sqrt2$: compressed ideal Bob observables are Hermitian contractions whose expectation converges to this value, the trace-norm optimisation over contractions equals that over involutions, and the Tsirelson bound \cite{Cirelson1980} supplies the matching upper bound. These arguments give neither a rate for the hard-bin convergence nor uniform control for occupations or amplitudes growing with $S$; Sec.~\ref{sec:finite} documents slow convergence over the computed range.

\section{Discussion and conclusions}
\label{sec:conclusion}
Whether a composite boson is bosonic depends on the detector. We have applied three detectors to one exactly solvable composite field and obtained three different answers, and we have traced the difference to one property of the composite: at finite Schmidt rank its quadrature has a discrete spectrum, the set of signed sums of the Schmidt amplitudes, which an elementary boson does not have.

A classical drive (A) is the fundamental harmonic of the quadrature and sees the composite through its commutator. Its Weyl defect is real-analytic in the Schmidt weights, is given by the product formula \eqref{eq:product}, has the dilute expansion \eqref{eq:expansion} in purity and third moment, and decays as $1/S$ for a Frenkel-like coboson on $S$ sites. Intensity correlations (C) see the state and not the quadrature; a quasi-Glauber state with $g^{(2)}=0.99999$ has a binned anticommutation defect of order one. Binned homodyne (B) resolves the lattice: its operators are piecewise constant in the weights, so that equal-moment spectra give different defects and different singlet probabilities; they are stable only within the margin of Proposition~\ref{prop:margin}, with lattice points on bin edges contributing the projectors of \eqref{eq:edgeAnti}; and on the Frenkel ladder they converge to the oscillator algebra so slowly that the defect is $0.37$ at $S=8000$, where the Weyl defect is $0.001$. The moment expansions of coboson theory, which control the drive, do not control the homodyne measurement; the harmonic defects of Appendix~\ref{app:harmonics} identify the mechanism, the loss of the fundamental-harmonic approximation at orders growing with $S$, but we have not established a rate.

For the task of extracting a qubit, the practical conclusion is that the imperfect involutions can be repaired. The two-involution reduction, applied on the spectral sector of the anticommutator below a cutoff, gives an exact Pauli algebra and an exact singlet projector at every rank, with the rejected weight and the measurement correction bounded by the defect. The success probability from two vacua factorises into the squared accepted weight and the mixedness of the filtered reduced qubit, Eq.~\eqref{eq:mixed}: filtering and the entanglement between the resolved qubit and its complementary logical factor play separate roles, the probability is nondecreasing in the cutoff, and the cutoff trades success probability against the admissible difference between the corrected observable and the binned quadrature. On the Frenkel ladder the construction converges strongly to that of Ref.~\cite{Sorella2026}, whose vacuum overlap is strictly positive at every scale (Proposition~\ref{prop:positive}), so the protocol succeeds for all sufficiently large flat-spectrum ranks. For Wannier-like spectra, which never become bosonic, finite-mode calculations (dense to Schmidt number about $5$, Lanczos beyond) give singlet probabilities up to $4\%$ for Schmidt numbers between $2$ and $9$, the loss being dominated by the vacuum weight on the one-dimensional Jordan blocks; a composite whose spectrum is so concentrated that $\sum_k\sqrt{\lambda_k}<\pi/2$ at $r=1$ -- an explicit condition on the amplitudes, which near-unit purity alone does not imply -- has $X=I$ and yields no qubit. For $z\le0.6$ the probabilities are stable at the percent level in the number of modes, and Proposition~\ref{prop:infinite} shows that the whole construction converges in the infinite-mode limit at all but countably many cutoffs; whether the cutoffs used are among the exceptions, and whether the limits are positive, is open.

\paragraph{Open questions and outlook.}
Four problems remain open. First, a hard-bin convergence rate on the Frenkel ladder requires control of the full square-wave anticommutator, including the edge projectors of \eqref{eq:edgeAnti}; the exponent $-0.15$ measured here is a finite-window value. Second, at fixed purity one must determine whether the chosen cutoffs are eigenvalues of $|K_\infty|$ and whether the limiting preparation probabilities are positive. Third, the step from model spectra to a specific system: a Schmidt spectrum derived from a confinement model of a quantum-dot exciton, or from a trapped pair, together with initial states of finite pair density, would say which composites carry a usable quadrature qubit and at what efficiency. Fourth, implementation requires coherent control of $\Zt$, which depends on both conjugate quadratures, and a heralded joint preparation; the self-testing bound \eqref{eq:selftest} can then certify the recovered algebra.

None of these is an obstacle to the main conclusion. A composite boson that the conventional diagnostics of bosonic behaviour place far from bosonic -- a normalisation ratio $F_2/F_1$ a third below unity, a Weyl defect of order one, a quadrature spectrum that is visibly discrete -- still yields, from its binned quadratures, an exact Pauli algebra and an exact singlet at the percent level, by a construction with an elementary-oscillator counterpart and with every loss computable from the Schmidt spectrum. At large flat rank the Weyl defect can already be small while binned quadratures still detect the finite spectral structure; near-unit intensity correlations likewise do not guarantee a small binned defect. Seen by the right detector, compositeness is not a defect that must be removed before a composite can serve as a qubit resource; it is a structure whose consequences for that use can be computed, and used.

\section*{Acknowledgments}
The author thanks S. P. Sorella for discussions, for confirming the values of Sec.~\ref{sec:limit}, and for informing him that a correction of Ref.~\cite{Sorella2026} has been made.

The author acknowledges the use of the LLM assistants Claude (Anthropic, model Fable 5.1) and OpenAI Codex (ChatGPT) for exploratory calculations, numerical cross-checks and language editing; all scientific content, calculations, interpretations and conclusions were checked by the author, who remains solely responsible for them. CNPq provided partial financial support through grant \mbox{319060/2025-0}.

\appendix

\section{Exact rank-four calculation and the raw sequential protocol}
\label{app:rank4}

Label the rank-four paired basis by bit strings $j=0,\dots,15$ and retain the even-parity integers $I=(0,3,5,6,9,10,12,15)$. Let $H_{jt}=(-1)^{\mathrm{popcount}(j\,\&\,t)}/4$, let $H_e$ contain these eight rows, and put $D_{jj}=i^{\,\mathrm{popcount}(j)}$. With $q_t=\sum_{k=0}^3(-1)^{t_k}\sqrt{\lambda_k}$, at $r=1$,
\begin{equation}
X=DH_e\,\mathrm{diag}[\sgn\cos q_t]\,H_e^{T}D^\dagger,\qquad Z=H_e\,\mathrm{diag}[\sgn\cos(\pi q_t/2)]\,H_e^{T}.
\end{equation}
All entries are rational once the bin signs are fixed; for the spectra \eqref{eq:pair} the smallest distance of a $q_t$ to a bin edge exceeds $0.02920$ and $0.19352$ respectively, so the signs are certified by rational square-root enclosures.

Since the uncorrected singlet polynomial need not be a projector, the raw protocol prepares the state obtained by the projective (L\"uders) measurement of the two-party observable $Z\otimes Z$ followed by that of $X\otimes X$, keeping the minus result each time -- joint projections onto the $-1$ eigenspaces, not local binned measurements followed by classical post-selection --
\begin{equation}
\ket{\psi_{\rm seq}}=\frac{\Pi^-_X\Pi^-_Z\ket{00}}{\sqrt{p_{\rm seq}}},\qquad \Pi^-_A=\frac{I-A\otimes A}2 ,
\end{equation}
the ordering being part of the protocol. For a real initial state $v$ the unnormalised coefficient matrix is
\begin{equation}
M=\frac14\big[vv^T-(Zv)(Zv)^T-(Xv)(Xv)^T+(XZv)(XZv)^T\big],\qquad p_{\rm seq}=\Tr(M^TM),\quad\Psi=M/\sqrt{p_{\rm seq}}.
\end{equation}
With Alice's settings fixed at $X,Z$ and the CHSH convention $\mathcal C=X\otimes(M_0+M_1)+Z\otimes(M_0-M_1)$ \cite{CHSH1969}, optimisation over Bob's dichotomic settings gives
\begin{equation}
\mathcal B_{\rm opt,B}=\|H_+\|_1+\|H_-\|_1,\qquad H_\pm=\Psi^T(X\pm Z)^T\Psi^*,
\end{equation}
since each term is a linear functional $\Tr(MH)$ maximised by $M=\sgn H$ with any dichotomic extension on its kernel. Exact evaluation gives
\begin{center}
\begin{tabular}{cccc}
\toprule
spectrum & $\varepsilon_0^2$ & $p_{\rm seq}$ & $\mathcal B_{\rm opt,B}$\\
\midrule
$a$ & $1$ & $3/32$ & $1+\sqrt3\simeq2.73205$\\
$b$ & $5/2$ & $23/128$ & $(2\sqrt{31}+27\sqrt2)/23\simeq2.14432$\\
\bottomrule
\end{tabular}
\end{center}
Each $H_\pm$ has six zero eigenvalues; the remaining pairs are $\{\mp1/2\}$, $\{\mp\sqrt3/2\}$ for $a$ and $\{(-3\mp2\sqrt{31})/46\}$, $\{(3\mp27\sqrt2)/46\}$ for $b$. These are protocol-dependent values, not intrinsic Bell violations attached to the spectra.

\section{Exact eight-mode construction}
\label{app:eight}

For $S=8$, $r=1$, in the even-number basis $(\ket0,\ket2,\ket4,\ket6,\ket8)$, the bin signs for the quadrature eigenvalue labels $j=0,\dots,8$ are $(-,-,+,+,+,+,+,-,-)$ for $X$ and $(-,-,-,+,+,+,-,-,-)$ for $Z$; no eigenvalue lies on an edge. The matrices are
\begingroup\small
\begin{equation}
X=\begin{pmatrix}
\frac{55}{64}&\frac{5\sqrt7}{32}&-\frac{\sqrt{70}}{64}&-\frac{3\sqrt7}{32}&\frac7{64}\\[2pt]
\frac{5\sqrt7}{32}&-\frac5{16}&\frac{7\sqrt{10}}{32}&\frac7{16}&-\frac{3\sqrt7}{32}\\[2pt]
-\frac{\sqrt{70}}{64}&\frac{7\sqrt{10}}{32}&-\frac3{32}&\frac{7\sqrt{10}}{32}&-\frac{\sqrt{70}}{64}\\[2pt]
-\frac{3\sqrt7}{32}&\frac7{16}&\frac{7\sqrt{10}}{32}&-\frac5{16}&\frac{5\sqrt7}{32}\\[2pt]
\frac7{64}&-\frac{3\sqrt7}{32}&-\frac{\sqrt{70}}{64}&\frac{5\sqrt7}{32}&\frac{55}{64}
\end{pmatrix},\quad
Z=\begin{pmatrix}
\frac{27}{64}&-\frac{9\sqrt7}{32}&\frac{3\sqrt{70}}{64}&-\frac{\sqrt7}{32}&-\frac{21}{64}\\[2pt]
-\frac{9\sqrt7}{32}&-\frac9{16}&-\frac{3\sqrt{10}}{32}&\frac3{16}&-\frac{\sqrt7}{32}\\[2pt]
\frac{3\sqrt{70}}{64}&-\frac{3\sqrt{10}}{32}&-\frac{23}{32}&-\frac{3\sqrt{10}}{32}&\frac{3\sqrt{70}}{64}\\[2pt]
-\frac{\sqrt7}{32}&\frac3{16}&-\frac{3\sqrt{10}}{32}&-\frac9{16}&-\frac{9\sqrt7}{32}\\[2pt]
-\frac{21}{64}&-\frac{\sqrt7}{32}&\frac{3\sqrt{70}}{64}&-\frac{9\sqrt7}{32}&\frac{27}{64}
\end{pmatrix}.
\end{equation}
\endgroup
The characteristic polynomial of $K=(XZ+ZX)/2$ is $(u+1)(8u-1)^2(32u+3)^2/65536$; with $\kappa=1/2$ the retained projectors and polar factor are
\begin{equation}
\begin{gathered}
E_a=\tfrac{256}{63}(K+I)(K+\tfrac{3}{32}I),\qquad E_b=-\tfrac{1024}{203}(K+I)(K-\tfrac18I),\qquad E=E_a+E_b,\\
J=\tfrac{8}{3\sqrt7}E_a+\tfrac{32}{\sqrt{1015}}E_b,\qquad \Zt=(Z-KX)J .
\end{gathered}
\end{equation}
Direct multiplication gives $E^2=E$, $\Zt^2=E$ and $\{EX,\Zt\}=0$. The vacuum entries of \eqref{eq:pv} are
\begin{equation}
t=\frac{225}{232},\qquad m_x=\frac{1539}{1856},\qquad m_y=0,\qquad m_z=\frac{7\sqrt{1015}}{1856}+\frac{\sqrt7}{8},
\end{equation}
giving
\begin{equation}
p_8(\ket0)=\frac{959-49\sqrt{145}}{29696}=0.0124246317901\ldots,\qquad \frac{|\vec m|}{t}=0.9732 .
\label{eq:p8}
\end{equation}
Since any $\kappa\in[1/8,1)$ retains the same even sector, \eqref{eq:p8} is independent of the cutoff in that range. The odd-parity $K$ eigenvalues are $-1,1,-3/4,-3/4$, so none is retained at $\kappa=1/2$; with the full normalisations of the $\alpha=\beta=1$ initial states,
\begin{equation}
p_8(G_1)=\frac{536363635311-37848641985\sqrt{145}}{293607926091776},\qquad
p_8(A_1)=\frac{361243454159-29996554273\sqrt{145}}{103543549572096},
\end{equation}
both positive, reproducing Table~\ref{tab:initial}. The symbolic multiplication of the two-party coefficient matrix $(uu^T-xx^T-yy^T-zz^T)/4$, with $u=Ev$, $x=X_Ev$, $y=\Yt v$, $z=\Zt v$, verifies all three singlet stabilisers.

\section{Field correlation versus pair-count correlation}
\label{app:g2}

For the normalised Anderson state $\ket{A_\beta}\propto e^{\beta B^\dagger}\ket0=\prod_k(1+\beta\sqrt{\lambda_k}\,d_k^\dagger)\ket0$ with flat spectrum, the pair-number law is binomial with parameter $q=|\beta|^2/(S+|\beta|^2)$, so $\langle(\hat N)_k\rangle=(S)_kq^k$ for falling factorials. Using \eqref{eq:intensity} and the two-step ladder action,
\begin{equation}
\begin{gathered}
I_A=\langle B^\dagger B\rangle=Sq-(S-1)q^2,\\
J_A=\langle B^{\dagger2}B^2\rangle=\frac{S-1}{S}\,q^2\big[S(S-1)-2(S-1)(S-2)q+(S-2)(S-3)q^2\big],
\end{gathered}
\end{equation}
the second from $(\hat N)_2(S-\hat N+1)(S-\hat N+2)=(\hat N)_4-2(S-1)(\hat N)_3+S(S-1)(\hat N)_2$. Hence, for $q>0$,
\begin{equation}
g^{(2)}_A=\frac{J_A}{I_A^2}=\frac{S-1}{S}\cdot\frac{S(S-1)-2(S-1)(S-2)q+(S-2)(S-3)q^2}{[S-(S-1)q]^2}.
\label{eq:g2A}
\end{equation}
This tends to $(1-1/S)^2$ as $q\to0$ and to the saturated Fock value $2(1-1/S)$ as $q\to1$, and it differs from the amplitude-independent value $\hat g_2=1-1/N_s$ given in Eq.~(53) of Ref.~\cite{Shiau2017} for the same state, Eq.~(50) there, with the field-operator definition of Eq.~(2) there. The value $1-1/N_s$ is exactly the binomial pair-count ratio $\langle\hat N(\hat N-1)\rangle/\langle\hat N\rangle^2$, which is distinct from the field correlation because of \eqref{eq:intensity}; we do not infer the origin of the discrepancy from this coincidence.

Two independent checks: (i) in the dilute limit $\beta\to0$, $B^{\dagger2}\ket0=\sqrt{2(1-1/S)}\ket2$ gives $J_A\simeq|\beta|^4(1-1/S)^2$ and $I_A^2\simeq|\beta|^4$, so $g^{(2)}_A\to(1-1/S)^2$ directly; (ii) for $S=2$, $\beta=1$ the normalised state is $(2/3,2/3,1/3)^T$, $B$ has unit first superdiagonal, $I_A=5/9$, $J_A=1/9$ and $g^{(2)}_A=9/25$ rather than $1/2$, confirmed by full two-site multiplication. In addition, \eqref{eq:g2A} was verified against direct evaluation of $\langle B^{\dagger2}B^2\rangle/\langle B^\dagger B\rangle^2$ on the normalised state $e^{\beta B^\dagger}\ket0$ in the full $2^S$-dimensional paired space for $S=3,4,5$ and $\beta=0.5,1,2$, to all digits; e.g.\ $S=3$, $\beta=1$ gives $8/15=0.5333$ against $2/3$. The state is that of Eq.~(50) of Ref.~\cite{Shiau2017}, the site product $\prod_p(1+\alpha N_s^{-1/2}e^{i\theta_p}\alpha_p^\dagger\beta_p^\dagger)\ket0$, and the correlation function is that of Eq.~(2) there. This correction does not affect the Frenkel ladder or the quasi-Glauber recurrence used in the main text; it does affect the statement in Ref.~\cite{Shiau2017} that the Anderson state has $\hat g_2<1$ for all $\alpha$, which by \eqref{eq:g2A} holds for $S\ge3$ only below a crossing density $q^*(S)$ with $g^{(2)}_A(q^*)=1$ ($q^*=0.61$ at $S=8$, $0.52$ at $S=32$); for $S=2$ it holds for all $\beta$.

\section{Harmonic-pair defects}
\label{app:harmonics}
For the flat spectrum, each ordered product of two local displacement factors maps the vacuum to a product of identical two-component vectors. Expanding each cosine into its two exponentials gives eight product states in the anticommutator. If their local vectors are $u_a$, then
\begin{equation}
\big\|\{\cos(uQ_r),\cos(vQ_w)\}\ket0\big\|^2
=\frac1{16}\sum_{a,b=1}^{8}\langle u_a|u_b\rangle^S.
\label{eq:harmProduct}
\end{equation}
This is independent of the diagonalisation of the Dicke ladder and agrees with it. Table~\ref{tab:harm} shows how the defect grows with the harmonic order at fixed rank: the displacement (fundamental) approximation is not uniform in the order, and at $S=4096$ the $(m,0)$ defect rises from $0.001$ at $m=0$ to $0.95$ at $m=8$. At $S=256$ the $Q_w$ grid is $q=k\pi/16$, so $\cos(17q)=(-1)^k\cos q$ and the $(0,8)$ entry equals the $(0,0)$ entry although the operators differ: aliasing is a further signature of the discrete spectrum.
\begin{table}[h]
\centering\small
\begin{tabular}{lrrr}
\toprule
$m$ & $n$ & $S=256$ & $S=4096$\\
\midrule
0 & 0 & 0.019289 & 0.001211\\
3 & 0 & 0.964957 & 0.093429\\
8 & 0 & 0.469799 & 0.947752\\
0 & 8 & 0.019289 & 0.213068\\
3 & 3 & 0.706428 & 0.853349\\
\bottomrule
\end{tabular}
\caption{Vacuum anticommutator defects $\|\{\cos(uQ_r),\cos(vQ_w)\}\ket0\|$ of harmonic pairs at $r=1$, $u=2m+1$, $v=2n+1$, from \eqref{eq:harmProduct}. In the oscillator every entry vanishes.}
\label{tab:harm}
\end{table}

\section{Numerical methods and verification}
\label{app:num}
Flat-spectrum quadratures are evaluated in the invariant Dicke ladder with a tridiagonal eigensolver; the bin signs are assigned on the analytic eigenvalue grid $(2j-S)/\sqrt S$, with a tie tolerance proportional to machine precision and both tie conventions computed at perfect-square ranks. Nonflat operators use normalised fast Hadamard transforms in the product bases of the local $\sigma_x$ and $\sigma_y$ operators, with the number-dependent phase for $X$; the implementation is tested against directly diagonalised tensor-product quadratures on complex vectors and reproduces the Dicke-ladder results for flat spectra to all digits. Dense recovery calculations use the invariant even vacuum sector ($2^{S-1}$ dimensions, used up to $S=14$) and record eigenpair residuals; larger calculations use a fully reorthogonalised Lanczos basis of $240$ steps started at the vacuum, whose accuracy was assessed by comparison with dense results ($S=13$ and $14$, Sec.~\ref{sec:purity}) and by varying the Krylov dimension (Table~\ref{tab:lanczos}). The Lanczos filter is accurate to about $10^{-4}$ when $\kappa$ is away from the vacuum spectral support of $K$ and to about $1$--$3\%$ when that support is quasi-continuous near $\kappa$; it does not reliably locate spectral atoms, which were therefore identified only from dense calculations. The cutoff curves of Fig.~\ref{fig:cutoff} use dense eigensystems and the computed eigenvalues of $|K|$ as step locations.

Two auxiliary tables are collected here. Table~\ref{tab:robust} gives the vacuum preparation probability at $S=8$ and $16$ under scale changes of $\pm10^{-8}$ with $b=\pi/(2r)$ enforced (Sec.~\ref{sec:finite}); Table~\ref{tab:lanczos} the dependence of the Lanczos filter on the Krylov dimension (Sec.~\ref{sec:purity}).
\begin{table}[h]
\centering\small
\begin{tabular}{lrrr}
\toprule
$S$ & $r$ & $p$ & $t$\\
\midrule
8 & $1-10^{-8}$ & 0.01242463 & 0.969828\\
8 & $1$ & 0.01242463 & 0.969828\\
8 & $1+10^{-8}$ & 0.01242463 & 0.969828\\
16 & $1-10^{-8}$ & 0.00513391 & 0.839637\\
16 & $1$ & 0.00002405 & 0.843640\\
16 & $1+10^{-8}$ & 0.00448050 & 0.810486\\
\bottomrule
\end{tabular}
\caption{Vacuum preparation at $\kappa=1/2$ under small changes of scale; exact edges use $F=+1$.}
\label{tab:robust}
\end{table}
\begin{table}[h]
\centering\small
\begin{tabular}{lrrrr}
\toprule
steps & $t_{0.4}$ & $p_{0.4}$ & $t_{0.6}$ & $p_{0.6}$\\
\midrule
240 & 0.531596 & 0.007379 & 0.737258 & 0.012558\\
480 & 0.537496 & 0.007503 & 0.739581 & 0.012625\\
960 & 0.543015 & 0.007610 & 0.737844 & 0.012605\\
\bottomrule
\end{tabular}
\caption{Dependence on the Krylov dimension at fixed $S=16$, $z=0.8$, $r=1$.}
\label{tab:lanczos}
\end{table}

To avoid cancellation in \eqref{eq:product}, put $u_k=\sin^2\alpha_k$, $v_k=\sin^2\beta_k$, and $\ell_k=4u_kv_k(u_k+v_k-2u_kv_k)$. Then $|g_k|^2=1-\ell_k$ and
\begin{equation}
\delta_D^2=-2\,\mathrm{expm1}(L_g)+4e^{L_g}\cos^2(\theta_g/2),\quad
L_g=\tfrac12\sum_k\log(1-\ell_k),\quad\theta_g=\sum_k\arg g_k,
\end{equation}
with $\delta_D=\sqrt2$ if a factor vanishes.

Assertion-based tests compare this expression with tensor-product exponentials, the fast transforms with direct matrices, dense and Lanczos small-rank results, the rank-four examples, the eight-mode probability and Pauli projectors, cutoff nesting, the reduced-qubit identity, the Anderson field correlation, the edge mass $127/512$ at $S=16$, and the smoothing bounds. The radicals of Appendices~\ref{app:rank4} and \ref{app:eight} come from exact symbolic computation. The flat scan through $S=8000$ and the nonflat scan through $19$ modes are archived with the code; the values of Tables~\ref{tab:scan}--\ref{tab:purity}, the rank-four and eight-mode quantities, the oscillator constants, the harmonic-pair defects and the Bell values at selected ranks were reproduced by independent implementations that do not share code with the production scripts (\texttt{code/independent/}); the remaining prose statistics (window minima, maxima and medians) are read from the archived scans. No experimental data are reported.

The scripts, the archived scans, the test and verification records and the sources of both versions of this paper are available in the repository \url{https://github.com/itzhr1/coboson-quadrature-spectrum}, archived at Zenodo under \href{https://doi.org/10.5281/zenodo.23960383}{doi:10.5281/zenodo.23960383} (which resolves to the latest version); \texttt{reproduce.sh} runs the tests, regenerates every table and figure block of the manuscripts from the archived data, and runs the independent checks.

\end{document}